\documentclass[
  aip,
  jcp,
  amsmath,amssymb,
  reprint
]{revtex4-1}

\usepackage[utf8]{inputenc}
\usepackage[T1]{fontenc}
\usepackage{graphicx}
\usepackage{dcolumn}
\usepackage{bm}
\usepackage{mathptmx}
\usepackage{etoolbox}
\usepackage{hyperref}
 
\newcommand{\includegraphicsmaybe}[2][]{%
  \IfFileExists{#2}{%
    \includegraphics[#1]{#2}%
  }{%
    \fbox{\parbox[c][0.22\textheight][c]{0.92\linewidth}{\centering\small Missing figure file:\\\texttt{#2}}}%
  }%
}

\makeatletter
\def\@email#1#2{%
  \endgroup
  \patchcmd{\titleblock@produce}
    {\frontmatter@RRAPformat}
    {\frontmatter@RRAPformat{\produce@RRAP{*#1\href{mailto:#2}{#2}}}\frontmatter@RRAPformat}
    {}{}%
}
\makeatother

\begin{document}

\preprint{Author-prepared version of the accepted manuscript; revised figures, results unchanged}
\preprint{Version of record: J. Chem. Phys. \textbf{164}, 214704 (2026); DOI: 10.1063/5.0336709}

\title{Fluctuation--Dissipation Framework for Size-Dependent Surface Tension}

\author{Sergii Burian}
\email{burian\_sergii@knu.ua}
\affiliation{Faculty of Physics, Taras Shevchenko National University of Kyiv, 64/13 Volodymyrska Street, Kyiv 01601, Ukraine}

\author{Yevhenii Shportun}
\affiliation{Faculty of Physics, Taras Shevchenko National University of Kyiv, 64/13 Volodymyrska Street, Kyiv 01601, Ukraine}

\author{Liudmyla Klochko}
\affiliation{Universit\'e de Lorraine, LORIA, 54000 Nancy, France}

\author{Leonid Bulavin}
\affiliation{Faculty of Physics, Taras Shevchenko National University of Kyiv, 64/13 Volodymyrska Street, Kyiv 01601, Ukraine}

\author{Dmytro Gavryushenko}
\affiliation{Faculty of Physics, Taras Shevchenko National University of Kyiv, 64/13 Volodymyrska Street, Kyiv 01601, Ukraine}

\author{Mykola Isaiev}
\affiliation{Universit\'e de Lorraine, CNRS, LEMTA, F-54000 Nancy, France}

\begin{abstract}
The size-dependent liquid-vapor surface tension controls phase change, wetting, and transport at nanoscales, yet its first curvature correction, the Tolman length, remains difficult to determine. We develop a thermodynamic and statistical-mechanical framework that relates this correction to bulk response properties of a one-component liquid near liquid-vapor coexistence. For curved interfaces, the analysis considers two local formulations of the same capillary-chemical balance, in excess pressures and in relative density deviations. For weakly compressible liquids in the regime emphasized here, the adopted asymmetric density-based formulation is the practically relevant one, with finite-curvature effects entering through vapor supersaturation under capillary equilibrium. At coexistence, the planar-limit value of the same Tolman length reduces to a combination of the liquid isothermal compressibility and its pressure derivative and can be recast as a bulk fluctuation-response observable of the homogeneous liquid in the isothermal-isobaric ensemble. In this representation, the planar-limit coefficient is determined by second and third central moments of the volume distribution, equivalently by the pressure response of the relative fluctuation width. For water, homogeneous \((N,P,T)\) simulations of SPC/E and TIP4P/2005 sample the bulk liquid, not an explicit liquid-vapor interface, and yield estimates near \(-0.7\)~\AA{} at \(300\)~K. An independent evaluation based on the IAPWS-IF97 industrial formulation gives \(-0.713\pm0.004\)~\AA{} at the same coexistence state and predicts a weakly nonmonotonic temperature dependence along coexistence. Beyond water, the framework applies to other one-component liquids in regimes where an accurate thermal equation of state or sufficiently converged bulk volume statistics are available.
\end{abstract}


\maketitle

\begin{center}
\small
\textit{Author-prepared version of the accepted manuscript; revised figures, results unchanged.
Version of record: J. Chem. Phys. \textbf{164}, 214704 (2026);
DOI: 10.1063/5.0336709.}
\end{center}

\section{Introduction}
\label{sec:intro}

At nanoscales, liquid--vapor interfacial free energies no longer scale trivially with area.
Because the interfacial region is a structured layer of finite thickness, changing its curvature reorganizes matter across that layer. As a result, the reversible work required to create interfacial area becomes intrinsically size dependent.
In nanoscale phase-change and interfacial transport phenomena, this dependence is therefore an effect of primary importance rather than a minor correction.
Curvature-dependent interfacial contributions arise in a broad range of settings where curvature is unavoidable, including phase behavior and related interfacial phenomena in confined and small-scale systems~\cite{Petters2020,Gallo2023,Cao2019}, cavitation and nucleation~\cite{Azouzi2013,Magaletti2021,Hajimirza2021}, and wetting, capillary transport, and multiphase processes on solids, in nanopores, and in hydrate-forming systems~\cite{Burian2024,Wang2016,Palodkar2017,Palodkar2018}.
A consistent description of this size dependence is therefore essential for a broad class of nanoscale processes.

The first curvature correction to the size dependence of the thermodynamic surface tension is commonly characterized by the Tolman length \(\delta_{\mathrm{Tolman}}\)~\cite{MalijevskyJackson2012,Sampayo2010}, yet its reliable determination remains difficult.
Most established routes require resolving an explicit interface and extracting sub-\AA{} curvature corrections from quantities that are highly sensitive to finite-size artifacts, dividing-surface conventions, and statistical convergence~\cite{Burian2024}.

A thermodynamic description of curved interfaces in heterogeneous one-component systems was developed by Gibbs~\cite{Gibbs1878OnTE}.
He introduced surface excesses (adsorptions) \(\Gamma\) and the method of dividing surfaces separating adjoining homogeneous phases \(\alpha\) and \(\beta\) with bulk densities \(\rho^\alpha\) and \(\rho^\beta\), thereby relating the curvature dependence of the surface tension \(\gamma\) to the ratio \(\Gamma/(\rho^\alpha-\rho^\beta)\).
In Gibbs' treatment, however, this ratio was not yet given an explicit geometric interpretation, and the resulting size dependence of \(\gamma\) remained implicit at that level.
For small spherical droplets, Tolman~\cite{10.1063/1.1747247} later identified this ratio with the geometric shift \(\delta_{\mathrm{Tolman}}:=R_e-R\) between the equimolar dividing surface of radius \(R_e\) and the surface of tension of radius \(R\), which led to the Gibbs--Tolman--K\"onig--Buff relation~\cite{PhysRevE.95.062801} for the size dependence of \(\gamma\).
Subsequent work generalized this result and showed that an analogous expression holds for cylindrical quasi-equilibrium interfaces in isotropic liquids, enabling analytical solutions for both spherical and cylindrical geometries~\cite{PhysRevE.95.062801}.
In these geometries, the Tolman approximation reads
\begin{equation}
	\label{eq:1.01}
	\gamma(R)=\frac{\gamma^\infty}{1+\frac{(n-1)\delta_{\mathrm{Tolman}}}{R}},
\end{equation}
where \(\gamma(R)\) is the thermodynamic surface tension at radius \(R\), \(\gamma^\infty\) is its planar limit as \(R\to\infty\), \(n\in\{2,3\}\) corresponds to cylindrical and spherical geometries, respectively, and \(\delta_{\mathrm{Tolman}}\) is the Tolman length.

Building on the Gibbs--Tolman framework, curvature corrections to \(\gamma\) have been analyzed using statistical-mechanical approaches~\cite{doi:10.1021/jp0108723,PhysRevE.105.015301}, alternative thermodynamic formalisms~\cite{10.1063/1.2167642,e21070670}, density-functional theories~\cite{10.1063/1.479650,10.1063/1.480434,10.1063/1.1354165}, and molecular-dynamics (MD) simulations~\cite{Joswiak2013,10.1063/1.3493464,Joswiak2016,Cui_2021}.
For a planar interface, Kirkwood and Buff~\cite{10.1063/1.1747248}, neglecting the vapor-phase contribution, obtained a statistical-mechanical expression involving the pair interaction potential and molecular distribution functions that is related to the planar-limit value of the Tolman length, although the required correlation functions are difficult to access for complex liquids.
Within the thermodynamic formalism of Kralchevsky and Nagayama~\cite{KRALCHEVSKY2000145,kralchevsky2001particles}, macroscopic interfacial parameters are written as integrals of local density and pressure-tensor profiles across and along the interface, enabling direct identification of the equimolar dividing surface and the surface of tension, and hence direct evaluation of \(\delta_{\mathrm{Tolman}}\).

Classical molecular-dynamics simulations allow one to compute components of the pressure tensor~\cite{10.1063/1.469505}, but accurate two-phase simulations with an explicitly resolved interface are computationally demanding and sensitive to finite-size effects and statistical convergence.
By contrast, bulk response coefficients, in particular the isothermal compressibility \(\beta_T\), can be obtained from volume fluctuations in homogeneous \((N,P,T)\) simulations or from a thermal equation of state.

Several relations of this type, expressed in terms of the isothermal compressibility, have been proposed previously~\cite{doi:10.1021/jp011028f,10.1063/1.2167642}.
Building on the work of Laaksonen and McGraw~\cite{A.Laaksonen_1996}, Bartell~\cite{doi:10.1021/jp011028f} assumed a linear dependence of the liquid molar volume on pressure, which led to a differential equation for \(\delta_{\mathrm{Tolman}}(R)\) and, in the planar limit, to
\begin{equation}
	\label{eq:1.02}
	\delta_{\mathrm{Tolman}}^{\mathrm{planar}}=-\,\gamma^\infty\beta_T,
\end{equation}
where \(\beta_T\equiv\beta_{T,l}\) is the isothermal compressibility of the liquid phase.
Wang and Zhu~\cite{Wang_2013} pointed out that this relation enforces \(\delta_{\mathrm{Tolman}}^{\mathrm{planar}}<0\) for any substance, which is inconsistent with some molecular-simulation results for simple fluids.

Blokhuis and Kuipers~\cite{10.1063/1.2167642} proposed a broader thermodynamic approach.
Along an isotherm near the coexistence curve, they considered a spherical liquid embryo of equimolar radius \(R_e\) in a metastable vapor, used the relation \(\mathrm{d}(\Delta P)=\Delta\rho\,\mathrm{d}\mu\), and expanded the Laplace pressure excess \(\Delta P\), the density difference \(\Delta\rho\) of the adjoining phases, and the chemical potential \(\mu\) in powers of \(1/R_e\) up to second order, neglecting the second-order curvature correction to \(\mu\) in order to close the expression.
In the planar limit this yields
\begin{equation}
	\label{eq:1.03}
	\delta_{\mathrm{Tolman}}^{\mathrm{planar}} \approx -\,\frac{\gamma^\infty}{(\Delta\rho_0)^2}
	\Bigl(\rho_{0,l}^2\,\beta_{T,l}-\rho_{0,v}^2\,\beta_{T,v}\Bigr),
\end{equation}
where \(\Delta\rho_0=\rho_{0,l}-\rho_{0,v}\) is the density difference at planar coexistence, and \(\beta_{T,\alpha}\) is the isothermal compressibility of phase \(\alpha\in\{l,v\}\).

Within a macroscopic thermodynamic formalism that uses the thermal equation of state \(P=f(V)\) at fixed \((T,N)\), our previous work~\cite{PhysRevE.95.062801} developed a capillary--compressibility construction for interfaces that are geometrically closed in one or more intrinsic directions. In such geometries, surface tension produces a net capillary contraction, while the liquid volume responds through its isothermal compressibility. For interfaces closed in one intrinsic direction (cylindrical geometry, \(n=2\)) or in two intrinsic directions (spherical geometry, \(n=3\)), one can obtain
\begin{equation}
	\label{eq:1.04}
	\delta_{\mathrm{Tolman}} \;\approx\; \frac{\gamma^\infty}{2}\left.
	\left[
	\frac{1}{\beta_{T,l}}\left(\frac{\partial \beta_{T,l}}{\partial P}\right)_{T}
	+ \frac{1}{n}\,\beta_{T,l}
	\right]\right|_{P_{\mathrm{sat}}(T)}.
\end{equation}
Here geometric closure means closure of the interface in the tangential directions along which surface tension generates net capillary contraction: two such directions for a sphere, one for a periodically continued cylinder, and none for a plane.
Unlike the simplified representation used in Ref.~\cite{PhysRevE.95.062801}, where the thermal-equation-of-state result was rewritten using the approximation \(f''(0)\approx -\,\beta_T'/\beta_T^{3}\), here we retain the exact identity \(f''(0)=1/\beta_T-\beta_T'/\beta_T^{3}\).
Accordingly, Eq.~\eqref{eq:1.04} is not the literal algebraic form written in Ref.~\cite{PhysRevE.95.062801}, but the corresponding form that follows from the same macroscopic construction when that simplifying step is not made.
Relation~\eqref{eq:1.04} makes clear that curvature corrections naturally involve both \(\beta_T\) and \(\beta_T'\), but it is derived for geometries with one or more closed intrinsic directions and therefore, by construction, does not provide a direct description for the planar interface.

Taken together, Eqs.~\eqref{eq:1.02}, \eqref{eq:1.03}, and~\eqref{eq:1.04} are all anchored in planar phase coexistence and relate the Tolman length, or its planar-limit value, to bulk thermodynamic response, but they treat curvature, size, and the vapor contribution in different ways.
Equation~\eqref{eq:1.02} relies on a linearization in \(\Delta P\) and depends only on the liquid compressibility, which guarantees a negative planar-limit Tolman length.
Equation~\eqref{eq:1.03} incorporates the liquid--vapor density contrast through the combinations \(\rho_{0,\alpha}^2\,\beta_{T,\alpha}\), but neglects second-order curvature corrections to \(\mu\).
Equation~\eqref{eq:1.04}, obtained for interfaces closed in one or more intrinsic directions, contains the explicit geometric factor \(n^{-1}\beta_{T,l}\) and retains explicitly the pressure derivative of the liquid compressibility, \(\beta'_{T,l}\). For water, this derivative contribution can be comparable to, or larger than, the term proportional to \(\beta_{T,l}\), and therefore cannot be neglected.

These considerations motivate the present work.
Our aim is to develop a thermodynamic and fluctuation-based framework in which the planar-limit value of the Tolman length is expressed in terms of bulk response properties of a one-component liquid near liquid--vapor coexistence.
The simulations used below are therefore homogeneous bulk simulations rather than direct interface-resolved calculations.
To this end, we examine two local formulations of the same capillary--chemical balance near planar coexistence: one in excess pressures and one in relative density deviations. These two formulations are controlled by different smallness assumptions. For weakly compressible liquids in the regime emphasized here, the formulation adopted in the present work is the asymmetric density-based one.

The remainder of the paper is organized as follows.
Section~\ref{sec:theory} presents the theoretical framework, including the coexistence reference state, capillary balance, two local expansions of the capillary--chemical equilibrium conditions, and the equivalent chemical-potential and fluctuation representations of the planar-limit result.
Section~\ref{sec:methods} describes the molecular models, simulation protocol, fluctuation estimators, block analysis, planar surface-tension input, and the IAPWS-IF97 implementation.
Section~\ref{sec:results} reports homogeneous \((N,P,T)\) benchmarks for SPC/E and TIP4P/2005 water and coexistence-curve results from IAPWS-IF97, including comparison with compressibility-only approximations.
Section~\ref{sec:discussion} discusses the physical interpretation, the relation to earlier approximations, and the scope of the framework, and Section~\ref{sec:conclusions} summarizes the main conclusions.
The Appendices collect the technical derivations, fluctuation identities, local expansion details, comparison variants, and water-specific hierarchy checks.

\section{Theoretical framework}
\label{sec:theory}

\subsection{Coexistence reference state and governing equations}
\label{sec:theory:setup}

Our aim is to relate the first curvature correction to the thermodynamic surface tension, and in particular the planar-limit value of the Tolman length, to bulk properties of a one-component liquid near liquid--vapor coexistence.
For a curved interface in thermal equilibrium, this problem is governed by two conditions that must hold simultaneously: mechanical equilibrium across the interface and chemical equilibrium between the two phases.

Using the unit normal directed from liquid to vapor, we introduce the signed mean curvature \(H\), defined on the surface of tension, so that the Laplace pressure jump is \(\Delta P \equiv P_l-P_v = H\,\gamma(R)\).
For the standard geometries, \(H=2/R\) for a spherical liquid droplet, \(H=1/R\) for a cylindrical liquid filament, and \(H=0\) for a plane; the corresponding vapor-bubble cases are obtained by the substitution \(H\to -H\).
Then, using the Tolman approximation introduced in Eq.~\eqref{eq:1.01} and written here in its first-order form, \(\gamma(R)\simeq \gamma^\infty\bigl(1-H\delta_{\mathrm{Tolman}}\bigr)\), we obtain
\begin{equation}
	\label{eq:2.01}
	\Delta P \equiv P_l-P_v
	=
	H\,\gamma(R)
	\simeq
	H\,\gamma^\infty\bigl(1-H\delta_{\mathrm{Tolman}}\bigr).
\end{equation}

Bulk thermodynamic response enters through the chemical-potential equality.
When the liquid and vapor chemical potentials are expanded near planar coexistence, the corresponding coefficients involve the densities, the isothermal compressibility, and the pressure derivative of the compressibility.
The corresponding local one-phase expansions are collected in Appendix~\ref{app:B}.
Thus, the curvature correction enters through mechanical equilibrium, while its relation to bulk thermodynamic response is obtained through chemical equilibrium.

As a reference state, we take planar liquid--vapor coexistence at temperature \(T\) and pressure \(P_0=P_{\mathrm{sat}}(T)\), where \(\mu_l(T,P_0)=\mu_v(T,P_0)\).
For a curved interface, we define the phase excess pressures
\begin{equation}
	\label{eq:2.02}
	\Delta P_v:=P_v-P_0,
	\qquad
	\Delta P_l:=P_l-P_0.
\end{equation}
Then \(\Delta P=\Delta P_l-\Delta P_v\), so that \(\Delta P_l=\Delta P_v+\Delta P\).
With the present sign convention, this corresponds to a liquid droplet or liquid filament surrounded by vapor, so that the liquid pressure exceeds the vapor pressure by the Laplace amount.
Mechanical and chemical equilibrium become
\begin{equation}
	\label{eq:2.03}
	\Delta P_l-\Delta P_v
	=
	H\,\gamma^\infty\bigl(1-H\delta_{\mathrm{Tolman}}\bigr),
\end{equation}
and
\begin{equation}
	\label{eq:2.04}
	\mu_v(P_0+\Delta P_v)
	=
	\mu_l(P_0+\Delta P_v+\Delta P).
\end{equation}

Equations~\eqref{eq:2.03} and~\eqref{eq:2.04} define a single capillary--chemical balance.
Near planar coexistence, this balance can be written in two different sets of local variables: directly in the excess pressures \(\Delta P_v\) and \(\Delta P_l\), or in the relative density deviations of the two phases.
The difference between these two formulations lies not in the equilibrium conditions themselves, which are the same, but in the smallness assumptions under which the corresponding expansions remain controlled.

\subsection{Formulation in excess pressures}
\label{sec:theory:pressure}

We first consider the chemical-potential equality~\eqref{eq:2.04} written in the excess pressures \(\Delta P_v\) and \(\Delta P_l\) about the planar coexistence state \((T,P_0)\).
In this formulation, \(\Delta P_v\) measures vapor supersaturation relative to planar coexistence, whereas \(\Delta P\) is the capillary pressure difference transmitted across the interface.
The vapor contribution is evaluated at the shifted argument \(P_0+\Delta P_v\), while the liquid contribution is evaluated at the shifted liquid argument \(P_0+\Delta P_v+\Delta P=P_0+\Delta P_l\).
Equivalently, on the liquid side one may regard the local expansion about \(P_0\) as written in the two increments \(\Delta P_v\) and \(\Delta P\), whose sum gives the total liquid excess pressure \(\Delta P_l\).
The corresponding local chemical-potential expansion in pressure variables is part of the one-phase input collected in Appendix~\ref{app:B}, while the treatment of the capillary--chemical balance in excess pressures is given in Appendix~\ref{app:D}.

At the retained order used here, the vapor contribution is kept only to first order in \(\Delta P_v\), whereas the liquid contribution is retained to second order at the shifted liquid argument \(P_0+\Delta P_v+\Delta P\).
This gives the lowest-order relation for the vapor supersaturation,
\begin{equation}
	\label{eq:2.05}
	\Delta P_v
	=
	H\,\gamma^\infty\,
	\frac{\rho_{v0}}{\rho_{l0}-\rho_{v0}}
	+
	O(H^2),
\end{equation}
which shows that \(\Delta P_v\) is linear in curvature.
In the dilute-vapor limit \(\rho_{v0}\ll\rho_{l0}\), Eq.~\eqref{eq:2.05} reduces to \(\Delta P_v\simeq H\gamma^\infty\,\rho_{v0}/\rho_{l0}\).

At the same retained order, the part of the balance containing \(\delta_{\mathrm{Tolman}}\) gives the planar-limit coefficient associated with this particular formulation in excess pressures,
\begin{equation}
	\label{eq:2.05b}
	\delta_{\mathrm{Tolman}}^{(\mathrm{pressure},\,\mathrm{planar})}
	=
	-\,\frac{\gamma^\infty}{2}\,\beta_{T,l}.
\end{equation}
This is the planar-limit coefficient obtained within the minimal retained formulation in excess pressures.
If one retains a stronger dependence on vapor supersaturation, one obtains finite-curvature expressions of a different logical status, discussed in Appendix~\ref{app:D}.

This formulation remains controlled when the excess pressures are sufficiently small for an expansion about \(P_0\) to be meaningful on the pressure scale set by planar coexistence.
This condition is logically distinct from weak compressibility: a liquid may be weakly compressible, while the Laplace overpressure is nevertheless too large for a controlled expansion in excess pressures.
For water near ambient coexistence, \(P_0=P_{\mathrm{sat}}(T)\) is only of the order of a few kPa, whereas the Laplace scale \(\gamma^\infty/R\) for nanometric radii reaches tens to hundreds of MPa.
In that regime, a formulation written directly in excess pressures is typically not controlled in practice.
This motivates the use of an alternative set of local variables that may remain small even when the excess pressures themselves do not.
For weakly compressible liquids, and in particular for water away from criticality, the relative density deviations provide such variables.

\subsection{Formulation in relative density deviations}
\label{sec:theory:density}

We therefore turn to the same capillary--chemical balance written in relative density deviations.
This formulation becomes natural when the excess pressures are no longer small on the coexistence-pressure scale, but the relative density response of the liquid remains small.
That is the relevant situation for weakly compressible liquids, for which a small relative density change provides a more useful local parameter than a small excess pressure.
For water away from criticality, this is the practically relevant case.
For each phase \(i\in\{l,v\}\), we write \(\Delta\rho_i \equiv (\rho_i-\rho_{i0})/\rho_{i0}\), where \(\rho_{i0}=\rho_i(T,P_0)\).
Along an isotherm and near the coexistence reference state, small relative density deviations are asymptotically equivalent to small values of \(\beta_{T,i}\Delta P_i\).
The local chemical-potential expansions in density variables are part of Appendix~\ref{app:B}, whereas the corresponding one-phase excess-pressure / relative-density series and their local inversion are collected in Appendix~\ref{app:C}.

In the derivation used here, the liquid and vapor chemical potentials are first expanded separately in the independent one-phase variables \(\Delta\rho_l\) and \(\Delta\rho_v\).
These relative density deviations are not linked by a separate mechanical-balance identity.
Their coupling is supplied only by chemical equilibrium.
Each branch is then re-expressed through its own excess pressure, namely \(\Delta P_l\) for the liquid and \(\Delta P_v\) for the vapor, and only afterward is the capillary relation used to write \(\Delta P_l=\Delta P_v+\Delta P\).
At the retained order, this procedure amounts to keeping the linear chemical coupling between \(\Delta\rho_l\) and \(\Delta\rho_v\), the second-order inverse pressure--density expansion on the liquid side, and only the first pressure response on the vapor side.
The corresponding derivation is given in Appendix~\ref{app:E}.
This yields
\begin{equation}
	\label{eq:2.06}
	\Delta P_v \frac{\rho_{l0}}{\rho_{v0}}
	=
	\Delta P_l
	+
	\frac{1}{2}\,\beta_{T,l}
	\left(
	1+\frac{\beta'_{T,l}}{\beta_{T,l}^{2}}
	\right)\Delta P_l^2,
\end{equation}
where \(\beta_{T,l}\) and \(\beta'_{T,l}\) are evaluated at the planar coexistence state \((T,P_0)\).
Here the liquid-side quantity is the total liquid excess pressure \(\Delta P_l=\Delta P_v+\Delta P\), not the capillary increment \(\Delta P\) taken separately.

Equation~\eqref{eq:2.06}, together with the capillary balance~\eqref{eq:2.03}, gives a mixed finite-curvature relation in which the vapor supersaturation remains explicit.
The lowest-order retained terms that do not contain \(\delta_{\mathrm{Tolman}}\) determine \(\Delta P_v\) to first order and reproduce Eq.~\eqref{eq:2.05}.
Within this retained expansion, the relation containing \(\delta_{\mathrm{Tolman}}\) is then extracted from the next retained curvature order, and only afterward is the planar limit taken.
This gives
\begin{equation}
	\label{eq:2.09}
	\delta_{\mathrm{Tolman}}^{\mathrm{planar}}
	=
	\frac{\gamma^\infty}{2}
	\left(
	\beta_{T,l}
	+
	\frac{\beta'_{T,l}}{\beta_{T,l}}
	\right).
\end{equation}

Here \(\delta_{\mathrm{Tolman}}^{\mathrm{planar}}\) is the planar-limit value of the Tolman length.
Equation~\eqref{eq:2.09} shows that this planar-limit value is determined by the liquid-state combination \(\beta_{T,l}+\beta'_{T,l}/\beta_{T,l}\), while finite-curvature effects enter through the vapor supersaturation that remains explicit before the planar limit is taken.
For weakly compressible liquids, and in particular for water away from criticality, this formulation remains useful over a substantially broader interval than the formulation written directly in excess pressures.
Water-specific support for the practical adequacy of the adopted mixed finite-curvature balance is given in Appendix~\ref{app:F}.

\subsection{Equivalent bulk forms of the planar-limit result}
\label{sec:theory:fdt}

Equation~\eqref{eq:2.09} expresses the planar-limit value of the Tolman length through the liquid response combination \(\beta_{T,l}+\beta'_{T,l}/\beta_{T,l}\) at planar coexistence.
This result can be rewritten in two equivalent bulk forms: through pressure derivatives of the liquid chemical potential, and through equilibrium volume fluctuations in the \((N,P,T)\) ensemble.

Using \(\mu'_l=(\partial\mu_l/\partial P)_T\), \(\mu''_l=(\partial^2\mu_l/\partial P^2)_T\), and \(\mu'''_l=(\partial^3\mu_l/\partial P^3)_T\), Eq.~\eqref{eq:2.09} becomes
\begin{equation}
	\label{eq:2.20}
	\delta_{\mathrm{Tolman}}^{\mathrm{planar}}(T)
	=
	\frac{\gamma^\infty}{2}
	\left(
	\frac{\mu'''_l}{\mu''_l}
	-
	2\,\frac{\mu''_l}{\mu'_l}
	\right)\Bigg|_{P_0}.
\end{equation}

To write the same result in fluctuation form, let \(\nu_1=\langle V\rangle\) be the mean volume, and let \(m_2=\langle(V-\langle V\rangle)^2\rangle\) and \(m_3=\langle(V-\langle V\rangle)^3\rangle\) be the second and third central moments of the \((N,P,T)\) volume distribution at \((T,P_0)\).
The isothermal compressibility is
\begin{equation}
	\label{eq:2.16}
	\beta_T
	=
	\frac{1}{k_B T}\,\frac{m_2}{\nu_1}\Bigg|_{P_0}.
\end{equation}
Using the pressure-derivative identities collected in Appendix~\ref{app:A}, Eq.~\eqref{eq:2.09} can be rewritten as
\begin{equation}
	\label{eq:2.21}
	\delta_{\mathrm{Tolman}}^{\mathrm{planar}}
	=
	\left.
	\frac{\gamma^\infty}{2k_B T}
	\left(
	2\,\frac{m_2}{\nu_1}
	-
	\frac{m_3}{m_2}
	\right)
	\right|_{P_0}.
\end{equation}
This form separates the contribution associated with the variance-to-mean ratio \(m_2/\nu_1\) from the first skewness-sensitive correction carried by \(m_3/m_2\).

Equivalently, writing \(\sigma_V=\sqrt{m_2}\), one obtains
\begin{equation}
	\label{eq:2.23}
	\frac{\delta_{\mathrm{Tolman}}^{\mathrm{planar}}}{\gamma^\infty}
	=
	\left.
	\left(\frac{\partial}{\partial P}\right)_{T,N}
	\ln\frac{\sigma_V}{\nu_1}
	\right|_{P_0}.
\end{equation}

Equations~\eqref{eq:2.20}, \eqref{eq:2.21}, and~\eqref{eq:2.23} are therefore equivalent forms of the same planar-limit result.

\subsection{Range of applicability and relation between the two formulations}
\label{sec:theory:applicability}

All quantities in the present analysis are evaluated in a neighborhood of the planar coexistence state \((T,P_0)\), where \(\Delta P_i=0\) and \(\Delta\rho_i=0\).
The two formulations discussed above are controlled by different smallness conditions.
The formulation in excess pressures is controlled when the excess pressures themselves remain small on the pressure scale set by planar coexistence.
The formulation in relative density deviations is controlled when the retained relative density changes remain small.
Along an isotherm, the latter condition is asymptotically equivalent to requiring the products \(\beta_{T,i}\Delta P_i\) to remain small on the corresponding retained branches.

For the regime emphasized in the present work, the practically relevant condition is the smallness of the liquid response, \(|\beta_{T,l}\Delta P_l|\ll1\).
This is why the formulation in relative density deviations is used here as the main derivation.
For weakly compressible water near ambient coexistence, the formulation in excess pressures about \(P_0=P_{\mathrm{sat}}(T)\) is usually not practically controlled in the nanometric regime, whereas the liquid relative-density response may remain small over a broader interval.
Closer to the critical region, where the coexistence pressure itself rises substantially, the formulation in excess pressures may again become practically useful.

The two formulations are therefore local treatments of the same capillary--chemical balance rather than competing physical models; neither is universally preferable, and regimes may exist in which neither smallness condition remains controlled.
\section{Methods}
\label{sec:methods}

\subsection{Molecular models and simulation protocol}
\label{sec:methods:models-md}

We considered two rigid, nonpolarizable water models, SPC/E and TIP4P/2005.
SPC/E is a three-site model with Lennard--Jones interactions acting on oxygen and partial charges located on oxygen and hydrogen sites~\cite{Berendsen1981,C1CP22168J}.
TIP4P/2005 is a four-site model with a massless charge site located on the HOH bisector and Lennard--Jones interactions acting only on oxygen~\cite{10.1063/1.2121687}.
All geometric constraints and force-field parameters follow the original parameterizations.
The corresponding numerical parameter sets are listed in the Supplementary Information (Supplementary Note~S1), while detailed LAMMPS implementation settings are summarized in Supplementary Note~S2.

All molecular-dynamics simulations were performed with LAMMPS (version 2~August~2023)~\cite{LAMMPS} in the isothermal--isobaric \((N,P,T)\) ensemble, using \texttt{real} units, three-dimensional periodic boundary conditions, and a time step of \(1\)~fs.
For each model, we simulated \(N=1000\) water molecules in a cubic periodic cell (Table~\ref{tab:md-setup}).
Nonbonded interactions comprised Lennard--Jones and Coulomb terms with model-specific real-space cutoffs, while long-range electrostatics were treated with PPPM summation.
Analytical Lennard--Jones tail corrections were applied to both energy and pressure.
Temperature and pressure were controlled with Nos\'e--Hoover thermostat and barostat chains at \(T=300\)~K and \(P_0=P_{\mathrm{sat}}(T)\) from IAPWS-IF97~\cite{Wagner2008}; at \(T=300\)~K this corresponds to \(P_0\simeq 3.5\)~kPa.
After an initial energy minimization and \((N,P,T)\) equilibration, we generated three statistically independent \(100\)~ns production trajectories for each model, corresponding to a total production time of \(300\)~ns per model.
Further protocol details, including model-specific settings and additional diagnostics, are given in Supplementary Note~S2.

\begin{table}[t]
\caption{\label{tab:md-setup}
Simulation settings for homogeneous-liquid \((N,P,T)\) benchmarks at \(T=300~\mathrm{K}\).
Shown are the key parameters for SPC/E and TIP4P/2005 simulations at coexistence pressure \(P_0=P_{\mathrm{sat}}(T)\).
Model-specific force-field values and detailed LAMMPS settings are given in Supplementary Notes~S1 and~S2.}
\begin{ruledtabular}
\begin{tabular}{lcc}
Parameter & SPC/E & TIP4P/2005 \\
\hline
Water molecules, \(N\) & 1000 & 1000 \\
Ensemble & \((N,P,T)\) & \((N,P,T)\) \\
Temperature, \(T\) (K) & 300 & 300 \\
Coexistence pressure, \(P_0\) (kPa) & 3.5 & 3.5 \\
Initial box edge (\AA) & 30 & 31 \\
Real-space cutoff (\AA) & 10 & 12 \\
Time step (fs) & 1 & 1 \\
Production per seed (ns) & 100 & 100 \\
Independent seeds & 3 & 3 \\
Reference block window, \(\tau_b\) (ns) & 50 & 50 \\
\end{tabular}
\end{ruledtabular}
\end{table}

During production, we recorded the instantaneous system volume \(V(t)\) from the LAMMPS thermodynamic output at every MD step.
Additional structural checks were performed to verify bulk-liquid stability; these diagnostics are reported in Supplementary Note~S2.

\subsection{Bulk fluctuation estimators and uncertainty analysis}
\label{sec:methods:estimators}

All bulk estimators were computed from the homogeneous-liquid \((N,P,T)\) volume time series \(V(t)\).
No interfacial configurations or pressure-tensor profiles are used in these estimates.
We use the mean volume \(\nu_1=\langle V\rangle\), together with the central moments \(m_2\) and \(m_3\), defined in Section~\ref{sec:theory:fdt}.
The isothermal compressibility then follows from Eq.~\eqref{eq:2.16}.

For the planar-limit value of the Tolman length, \(\delta_{\mathrm{Tolman}}^{\mathrm{planar}}\), we first compute the \(\gamma^\infty\)-free estimator
\begin{equation}
\begin{split}
X
&\equiv
\frac{\delta_{\mathrm{Tolman}}^{\mathrm{planar}}}{\gamma^\infty}
\\
&=
\frac{1}{2k_B T}
\left(
2\,\frac{m_2}{\nu_1}
-
\frac{m_3}{m_2}
\right),
\end{split}
\label{eq:methods:X}
\end{equation}
which is equivalent to Eq.~\eqref{eq:2.21} divided by \(\gamma^\infty\) and, in response form, to Eq.~\eqref{eq:2.23}.
The corresponding dimensional value is then reported as
\begin{equation}
\delta_{\mathrm{Tolman}}^{\mathrm{planar}}=\gamma^\infty X,
\label{eq:methods:delta-from-X}
\end{equation}
with \(\gamma^\infty\) supplied independently as the planar surface-tension input (Section~\ref{sec:methods:gamma}).

To quantify convergence and the dependence on sampling timescale, we use a block-averaging protocol.
Each \(100\)~ns trajectory is partitioned into contiguous nonoverlapping blocks of duration \(\tau_b\), giving \(N_b=\lfloor 100\,\mathrm{ns}/\tau_b\rfloor\) blocks per seed.
Within each block \(b\), we compute \((\nu_1^{(b)},m_2^{(b)},m_3^{(b)})\) and the corresponding block estimates \(\beta_T^{(b)}\) and \(X^{(b)}\).
For each \(\tau_b\), we first average over blocks within each seed and then average across the three independent seeds; the seeds are analyzed separately and are never concatenated.
Unless stated otherwise, the reference estimate corresponds to \(\tau_b=50\)~ns.
This choice already lies in the plateau regime of the block-size diagnostics while still retaining two blocks per seed, so that a within-seed contribution to the uncertainty estimate remains available.
The full \(\tau_b\)-dependence, together with time-shuffled controls and autocorrelation-based diagnostics of the block plateau, is reported in the Supplementary Information (Supplementary Fig.~S1 and Supplementary Note~S2).

Uncertainties for seed-averaged estimators are reported as \(1\sigma\) standard uncertainties combining within-seed block statistics, when \(N_b\ge 2\), and between-seed variability.
For an estimator \(Y\in\{\beta_T,X\}\), we define
\begin{equation}
u_Y^2(\tau_b)
=
\mathrm{SEM}_{\mathrm{between}}^2(\tau_b)
+
\overline{\mathrm{SEM}}_{\mathrm{within}}^{\,2}(\tau_b),
\label{eq:methods:uY}
\end{equation}
where \(\mathrm{SEM}_{\mathrm{between}}=s_{\mathrm{between}}/\sqrt{3}\) is obtained from the standard deviation of the three seed-level estimates, and \(\overline{\mathrm{SEM}}_{\mathrm{within}}\) is the mean, over seeds, of the within-seed standard error of the block mean, \(\mathrm{SEM}_{\mathrm{within}}=s_{\mathrm{within}}/\sqrt{N_b}\).
For block windows such that \(N_b<2\) per seed, only the between-seed contribution is available.
Because \(X\) depends explicitly on the third central moment \(m_3\), its sampling uncertainty is generally larger than that of \(\beta_T\); this point is further documented by the block-size, autocorrelation, and shuffled diagnostics in Supplementary Note~S2.

\subsection{Planar surface-tension input}
\label{sec:methods:gamma}

\paragraph{Water-model surface tension at \(300\)~K.}
For SPC/E and TIP4P/2005 at \(T=300\)~K, we use planar liquid--vapor surface-tension values from Vega and de Miguel~\cite{VegaDeMiguel2007}: for SPC/E, \(\gamma^\infty=63.6\)~mN\,m\(^{-1}\) with reported uncertainty \(\sigma_\gamma=1.5\)~mN\,m\(^{-1}\), and for TIP4P/2005, \(\gamma^\infty=69.3\)~mN\,m\(^{-1}\) with \(\sigma_\gamma=0.9\)~mN\,m\(^{-1}\).
These values are external inputs and are not fitted to the present bulk fluctuation analysis; they enter only at the reporting stage through Eq.~\eqref{eq:methods:delta-from-X}.
For the force-field benchmarks, they are treated as model-specific planar surface-tension inputs, so that the dimensional estimate \(\delta_{\mathrm{Tolman}}^{\mathrm{planar}}=\gamma^\infty X\) remains consistent with the same molecular model from which \(X\) is obtained.

\paragraph{Real-water surface tension along coexistence.}
For IF97-based coexistence curves and comparison with compressibility-only relations, we require \(\gamma^\infty(T)\) along liquid--vapor coexistence.
We obtain \(\gamma^\infty(T)\) from the IAPWS surface-tension correlation along saturation.
Its finite accuracy is treated as an input tolerance in the deterministic propagation described in Supplementary Note~S3.

Uncertainty propagation for \(\delta_{\mathrm{Tolman}}^{\mathrm{planar}}=\gamma^\infty X\) in the MD-based estimates is carried out by first-order propagation, treating \(\gamma^\infty\) and \(X\) as independent inputs:
\begin{equation}
u_\delta^2=(\gamma^\infty u_X)^2+(X\,\sigma_\gamma)^2,
\label{eq:methods:udelta}
\end{equation}
where \(u_X\) is the sampling uncertainty from the block and seed analysis.

\subsection{IAPWS-IF97 implementation and uncertainty bands}
\label{sec:methods:if97}

Thermodynamic reference quantities for real water are obtained from the IAPWS-IF97 industrial formulation~\cite{Wagner2008}.
Along coexistence, we evaluate the saturation pressure \(P_{\mathrm{sat}}(T)\) using IF97 Region~4, saturated-liquid properties using IF97 Region~1 at \((T,P_0)\) with \(P_0=P_{\mathrm{sat}}(T)\), and vapor properties, where needed, using IF97 Region~2 at the same coexistence state.

To compute \(\delta_{\mathrm{Tolman}}^{\mathrm{planar}}(T)\) from Eq.~\eqref{eq:2.20}, we evaluate the saturated-liquid specific Gibbs free energy \(g(T,P)\), which for a pure substance is proportional to the chemical potential, together with its pressure derivatives at fixed \(T\) by analytic differentiation of the IF97 functional forms.
Because Eq.~\eqref{eq:2.20} involves only ratios of pressure derivatives, this constant proportionality does not affect the reported result.
No numerical differentiation with respect to \(P\) is used in the reported evaluation.
Internal checks between algebraically equivalent representations were used to validate the implementation; details are given in Supplementary Note~S3.
For comparison with compressibility-only approximations, the Bartell and Blokhuis--Kuipers relations were evaluated from the same IAPWS-IF97 coexistence data and the same \(\gamma^\infty(T)\) input as the full estimator.

For the IF97-based coexistence curve, the shaded bands shown in the Results represent \emph{input-accuracy (tolerance)} effects associated with the finite accuracy of empirical correlations, specifically IF97 and \(\gamma^\infty(T)\), rather than statistical sampling uncertainty.
We report both (i) a \(k=1\) standard-uncertainty band obtained by mapping each tolerance half-width \(\Delta x\) to \(u(x)=\Delta x/\sqrt{3}\) and propagating by linearization, and (ii) a conservative tolerance envelope obtained by shifting inputs by \(\pm\Delta x\) and recomputing the curve.
The deterministic propagation protocol is documented in Supplementary Note~S3.
These bands should therefore be interpreted as deterministic finite-accuracy representations of empirical inputs, not as confidence intervals and not as measures of molecular-dynamics sampling uncertainty.

\subsection{Data processing and reproducibility}
\label{sec:methods:repro}

Trajectory post-processing and statistical analysis were performed in Python using in-house scripts based on NumPy and SciPy.
Raw LAMMPS thermodynamic output was converted into analysis-ready time series and processed with the block-averaging protocol described above.
Intermediate results used for plotting and tabulation were cached to ensure reproducible regeneration from the same inputs.
LAMMPS input scripts and analysis codes are available from the corresponding author upon reasonable request.
Additional simulation details, diagnostics, and uncertainty-propagation procedures are provided in the Supplementary Information.
\section{Results}
\label{sec:results}

\subsection{Bulk \((N,P,T)\) benchmark at \(T=300\,\mathrm{K}\)}
\label{sec:results:md300}

We first assess the bulk fluctuation--response estimator for the planar-limit value of the Tolman length, \(\delta_{\mathrm{Tolman}}^{\mathrm{planar}}\), at liquid--vapor coexistence using homogeneous \((N,P,T)\) simulations of SPC/E and TIP4P/2005 water.
The estimator is evaluated from the moment representation and its equivalent response form, Eqs.~\eqref{eq:2.21} and~\eqref{eq:2.23}.
All simulations were performed at \(T=300\,\mathrm{K}\) and \(P_0=P_{\mathrm{sat}}(T)\) from IAPWS-IF97, with \(P_0\simeq 3.5\)~kPa, using three statistically independent \(100\)~ns production runs per model (Methods, Table~\ref{tab:md-setup}).

From each trajectory, we estimate \(\beta_T\) and the \(\gamma^\infty\)-free quantity \(X\equiv \delta_{\mathrm{Tolman}}^{\mathrm{planar}}/\gamma^\infty\) using the volume-moment estimators described in Section~\ref{sec:methods:estimators}, and then report \(\delta_{\mathrm{Tolman}}^{\mathrm{planar}}=\gamma^\infty X\), with \(\gamma^\infty(300\,\mathrm{K})\) taken from Ref.~\cite{VegaDeMiguel2007}.
Unless stated otherwise, the reported point estimates correspond to the reference block window \(\tau_b=50\)~ns.
Extended block-window diagnostics and original-versus-shuffled controls are reported in the Supplementary Information (Supplementary Fig.~S1 and Table~S2).
The MD uncertainties quoted below are \(1\sigma\) standard uncertainties; for \(\delta_{\mathrm{Tolman}}^{\mathrm{planar}}=\gamma^\infty X\), they include propagation of the external uncertainty in \(\gamma^\infty\), as described in Methods.

Table~\ref{tab:results} summarizes the resulting bulk estimates.
At \(300\,\mathrm{K}\), we obtain \(\beta_T\simeq 0.455\,\mathrm{GPa}^{-1}\) and \(\delta_{\mathrm{Tolman}}^{\mathrm{planar}}\simeq -0.69\,\text{\AA}\) for SPC/E, and \(\beta_T\simeq 0.471\,\mathrm{GPa}^{-1}\) and \(\delta_{\mathrm{Tolman}}^{\mathrm{planar}}\simeq -0.66\,\text{\AA}\) for TIP4P/2005.
Within uncertainties, both models therefore yield mutually consistent negative planar-limit values of sub-\AA{} magnitude.

For comparison, Table~\ref{tab:results} also includes an independent evaluation based on the IAPWS-IF97 industrial formulation at the same coexistence state \((T,P_0)\), obtained from the chemical-potential derivative representation, Eq.~\eqref{eq:2.20}, implemented through analytic pressure derivatives of the saturated-liquid specific Gibbs free energy (Methods).
The IF97-based estimate gives \(\delta_{\mathrm{Tolman}}^{\mathrm{planar}}=-0.713\pm0.004\,\text{\AA}\) at the same coexistence state.
This agreement should be interpreted narrowly.
The benchmark is intended to test the internal consistency and practical usability of the bulk fluctuation--response construction, not to establish the absolute accuracy of a particular water model or of the external planar surface-tension input \(\gamma^\infty\).
Within that scope, the comparison shows that, once the skewness-sensitive bulk-response contribution, encoded thermodynamically in \(\beta'_{T,l}\) and in the fluctuation representation through the \(m_3/m_2\) term, is retained, the planar-limit value \(\delta_{\mathrm{Tolman}}^{\mathrm{planar}}\) can be inferred from bulk \((N,P,T)\) volume statistics without resolving an explicit liquid--vapor interface.

\begin{table}[t]
\caption{\label{tab:results}
Bulk \((N,P,T)\) estimates at \(T=300\,\mathrm{K}\).
Shown are the isothermal compressibility \(\beta_T\) and the planar-limit value \(\delta_{\mathrm{Tolman}}^{\mathrm{planar}}\) at coexistence, obtained from homogeneous \((N,P,T)\) simulations of SPC/E and TIP4P/2005 at \(P_0=P_{\mathrm{sat}}(T)\) (\(P_0\simeq 3.5\)~kPa).
Three independent \(100\)~ns production trajectories were analyzed without concatenation.
The block window \(\tau_b\) controls the coarse-graining used to estimate the \((N,P,T)\) volume moments.
The reported values of \(\delta_{\mathrm{Tolman}}^{\mathrm{planar}}\) were obtained as \(\delta_{\mathrm{Tolman}}^{\mathrm{planar}}=\gamma^\infty X\), where \(X\) is the direct bulk fluctuation estimator defined in Eq.~\eqref{eq:methods:X}; \(\gamma^\infty(300\,\mathrm{K})\) is taken from Ref.~\cite{VegaDeMiguel2007} and enters only as an external planar surface-tension input. The quoted MD uncertainties are \(1\sigma\) and include propagation of the input uncertainty in \(\gamma^\infty\).
For \(\tau_b=100\)~ns, each seed contributes a single block, so the within-seed contribution is no longer available; the smaller quoted uncertainty at this block window should therefore not be interpreted as evidence of a strictly more precise estimate.
The IF97 reference uses Eq.~\protect\eqref{eq:2.20} evaluated at the same \((T,P_0)\); its quoted uncertainty reflects input-accuracy propagation rather than sampling uncertainty.}
\begin{ruledtabular}
\begin{tabular}{lccc}
Model / reference & \(\tau_b\) (ns) & \(\beta_T\) (GPa\(^{-1}\)) & \(\delta_{\mathrm{Tolman}}^{\mathrm{planar}}\) (\AA) \\
\hline
SPC/E & 50  & \(0.455 \pm 0.001\) & \(-0.69 \pm 0.07\) \\
SPC/E & 100 & \(0.456 \pm 0.001\) & \(-0.69 \pm 0.03\) \\
TIP4P/2005 & 50  & \(0.471 \pm 0.001\) & \(-0.66 \pm 0.06\) \\
TIP4P/2005 & 100 & \(0.472 \pm 0.001\) & \(-0.66 \pm 0.04\) \\
IAPWS-IF97 + Eq.~\protect\eqref{eq:2.20} & -- & \(0.4497 \pm 0.0002\) & \(-0.713 \pm 0.004\) \\
\end{tabular}
\end{ruledtabular}
\end{table}

\subsection{Convergence and diagnostics of the bulk estimators}
\label{sec:results:diagnostics}

\begin{figure*}[t]
	\centering
	\includegraphicsmaybe[width=\textwidth]{\detokenize{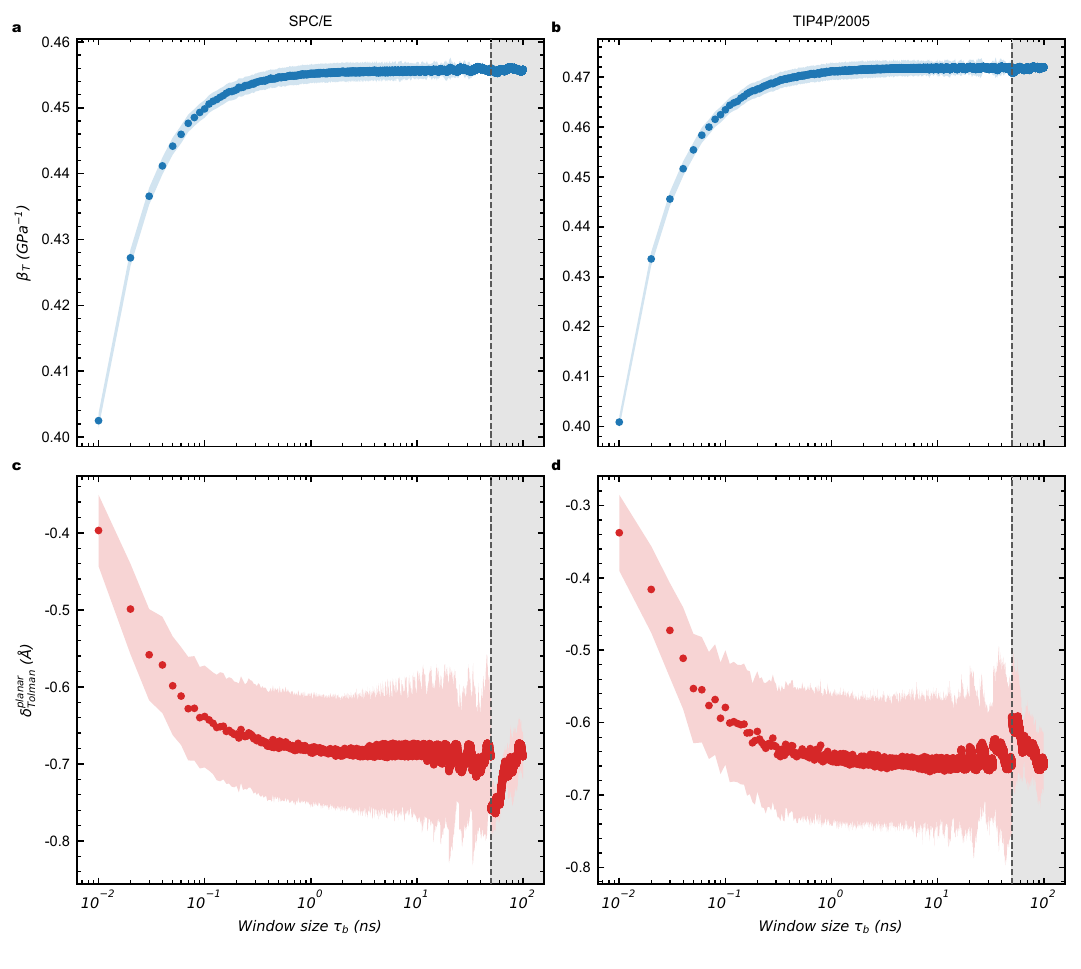}}
	\caption{\label{fig:block-analysis}
	Block-size dependence of bulk-fluctuation estimators at \(T=300\,\mathrm{K}\).
	Liquid isothermal compressibility \(\beta_T\) [(a),(b)] and the planar-limit value \(\delta_{\mathrm{Tolman}}^{\mathrm{planar}}\) [(c),(d)], estimated from homogeneous \((N,P,T)\) simulations of SPC/E (left) and TIP4P/2005 (right) water at \(P_0=P_{\mathrm{sat}}(T)\) from IAPWS-IF97.
	For each block window \(\tau_b\), each trajectory was partitioned into contiguous nonoverlapping blocks; within each block we computed \(\nu_1\), \(m_2\), and \(m_3\), and mapped them to \(\beta_T\) and \(X\equiv\delta_{\mathrm{Tolman}}^{\mathrm{planar}}/\gamma^\infty\) via Eqs.~\protect\eqref{eq:2.16} and~\protect\eqref{eq:methods:X}.
	Symbols show the mean over three independent seeds; shaded bands indicate \(\pm 1\sigma\) uncertainties constructed from within-seed and between-seed variability.
	The reported values of \(\delta_{\mathrm{Tolman}}^{\mathrm{planar}}\) are obtained as \(\delta_{\mathrm{Tolman}}^{\mathrm{planar}}=\gamma^\infty X\), using \(\gamma^\infty(300\,\mathrm{K})\) from Ref.~\cite{VegaDeMiguel2007}; the corresponding input uncertainty in \(\gamma^\infty\) is propagated in quadrature into the \(\delta_{\mathrm{Tolman}}^{\mathrm{planar}}\) bands.
	The vertical dashed line marks \(\tau_b=50\,\mathrm{ns}\).
	The gray region indicates \(\tau_b>50\,\mathrm{ns}\), where \(N_b<2\) per seed and only between-seed variability contributes to the uncertainty band.}
\end{figure*}

To assess the practical convergence of the bulk estimators, Fig.~\ref{fig:block-analysis} shows the block-size dependence of \(\beta_T\) and \(\delta_{\mathrm{Tolman}}^{\mathrm{planar}}\) for SPC/E and TIP4P/2005 at \(300\,\mathrm{K}\), based on three independent \(100\)~ns trajectories analyzed without concatenation.
Across the explored block windows, \(\beta_T\) remains stable within uncertainty, whereas \(\delta_{\mathrm{Tolman}}^{\mathrm{planar}}\) exhibits broader uncertainty bands.
This difference is expected because the planar-limit estimator depends explicitly on \(m_3\), which is more sensitive to tail statistics, and hence to the effective sample size, than the compressibility estimator.

The reference estimates quoted in Table~\ref{tab:results} correspond to \(\tau_b=50\)~ns.
This choice lies in the plateau regime while still retaining two blocks per seed; at \(\tau_b=100\)~ns, by contrast, each seed contributes only a single block.
A direct original-versus-shuffled comparison is provided in the Supplementary Information (Supplementary Fig.~S1 and Table~S2): the disappearance of the short-\(\tau_b\) drift after shuffling shows that the finite-window bias at small \(\tau_b\) originates from temporal correlations in the MD dynamics rather than from the block-averaging procedure itself.

\subsection{Temperature dependence along coexistence from IAPWS-IF97 and comparison with compressibility-only relations}
\label{sec:results:if97}

\begin{figure*}[t]
	\centering
	\includegraphicsmaybe[width=\textwidth]{\detokenize{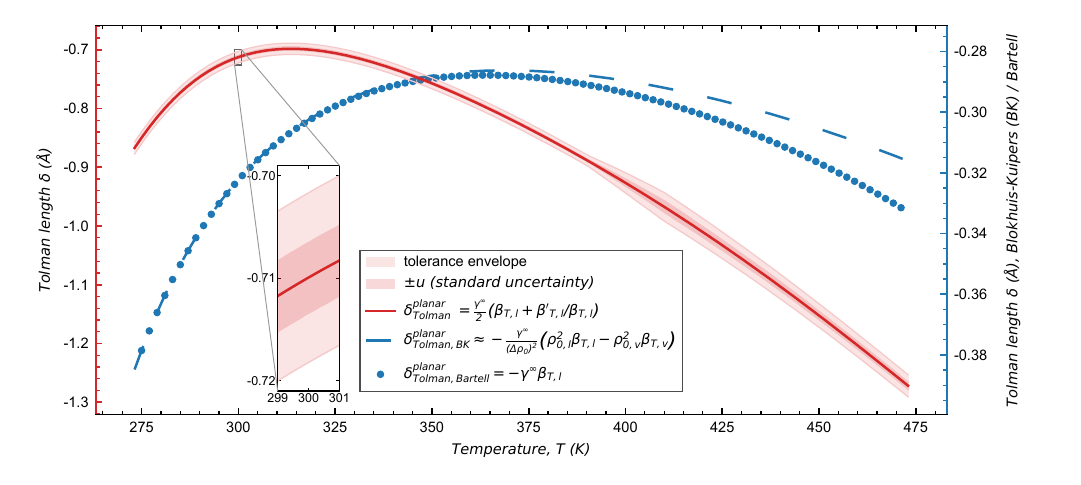}}
	\caption{\label{fig:Tolman_IF97}
	Planar-limit value \(\delta_{\mathrm{Tolman}}^{\mathrm{planar}}\) along coexistence from IAPWS-IF97.
	Temperature dependence of \(\delta_{\mathrm{Tolman}}^{\mathrm{planar}}(T)\) evaluated on coexistence for \(T=273\text{--}473\,\mathrm{K}\) at \(P=P_{\mathrm{sat}}(T)\).
	The red curve is obtained from Eq.~\protect\eqref{eq:2.20} using analytic pressure derivatives of the saturated-liquid specific Gibbs free energy from IF97 Region~1 and the saturation-pressure relation from IF97 Region~4.
	The blue dashed curve shows the Blokhuis--Kuipers relation, while the blue symbols show the Bartell relation, Eqs.~\protect\eqref{eq:1.03} and~\protect\eqref{eq:1.02}, evaluated at the same coexistence state and using the same \(\gamma^\infty(T)\) input (Methods; Supplementary Note~S3).
	Shaded regions quantify input-accuracy effects associated with IF97 and \(\gamma^\infty(T)\); the inner band represents the \(k=1\) standard-uncertainty propagation, while the outer envelope corresponds to a conservative tolerance representation.
	These bands reflect deterministic input tolerances of empirical correlations rather than statistical sampling uncertainty.}
\end{figure*}

Having established the bulk estimator at \(300\,\mathrm{K}\) and its convergence behavior, we next evaluate \(\delta_{\mathrm{Tolman}}^{\mathrm{planar}}(T)\) along liquid--vapor coexistence using IAPWS-IF97.
The red curve in Fig.~\ref{fig:Tolman_IF97} is obtained from the chemical-potential derivative representation, Eq.~\eqref{eq:2.20}, evaluated at \((T,P_0)\) with \(P_0=P_{\mathrm{sat}}(T)\), using analytic pressure derivatives of the saturated-liquid specific Gibbs free energy, as described in Methods.

For comparison, Fig.~\ref{fig:Tolman_IF97} also shows the Bartell and Blokhuis--Kuipers compressibility-based approximations, Eqs.~\eqref{eq:1.02} and~\eqref{eq:1.03}, evaluated at the same coexistence state and using the same planar surface-tension input \(\gamma^\infty(T)\) (Methods; Supplementary Note~S3).
Differences between the curves therefore reflect the thermodynamic structure of the estimators themselves rather than differences in coexistence input or surface-tension data.

Across \(T=273\text{--}473\,\mathrm{K}\), all estimators predict \(\delta_{\mathrm{Tolman}}^{\mathrm{planar}}<0\).
However, retaining the full liquid-state response combination \(\beta_{T,l}+\beta'_{T,l}/\beta_{T,l}\) yields substantially larger magnitudes and a markedly nonlinear, weakly nonmonotonic temperature dependence, whereas compressibility-only relations produce smaller magnitudes and smoother trends.
The systematic offset between the full IF97-based estimator and the Bartell and Blokhuis--Kuipers curves therefore quantifies the contribution associated with \(\beta'_{T,l}\), or equivalently with the \(m_3/m_2\) sector in the fluctuation representation.
In this sense, approximations that retain only \(\beta_T\) keep only the variance-controlled part of the bulk response, whereas the planar-limit result remains sensitive to the skewness-sensitive \(m_3/m_2\) sector of the \((N,P,T)\) volume distribution.

\section{Discussion}
\label{sec:discussion}

The central physical result of the present work is that the planar-limit value of the Tolman length at liquid--vapor coexistence is determined by liquid-state response properties evaluated at the coexistence reference state \((T,P_0)\).
In the formulation adopted in the manuscript, Eq.~\eqref{eq:2.09} shows that \(\delta_{\mathrm{Tolman}}^{\mathrm{planar}}\) depends on the liquid isothermal compressibility together with its pressure derivative.
Accordingly, the first curvature correction to the thermodynamic surface tension can, in the planar limit, be inferred from response properties of the homogeneous liquid.
This changes the practical statement of the problem: instead of extracting a sub-\AA{} correction from explicit two-phase configurations, one may determine the same planar-limit quantity from bulk thermodynamic response, provided the local near-coexistence expansion remains controlled.

This result is not introduced as an independent bulk assumption.
It follows from the simultaneous use of the two equilibrium conditions that govern a curved interface near planar coexistence.
Mechanical equilibrium introduces the curvature correction through the curvature dependence of the thermodynamic surface tension, while chemical equilibrium transfers this curvature contribution into the pressure response of the adjoining phases through the chemical potentials.
When these two conditions are expanded consistently near the planar coexistence state, the planar-limit coefficient becomes expressible through liquid-state response characteristics.

The analysis also clarifies the status of the two local formulations considered in the manuscript.
One formulation expands the capillary--chemical balance in the excess pressures of the two phases.
The other expands the same balance in their relative density deviations.
These are not different physical models.
They are two local formulations of the same equilibrium problem, controlled by different smallness conditions and therefore useful in different thermodynamic regimes.

For weakly compressible water in the nanometric regime emphasized here, the practically useful formulation is the one written in relative density deviations.
The reason is straightforward.
Near ambient coexistence, the reference pressure \(P_0=P_{\mathrm{sat}}(T)\) is only of the order of a few kPa, whereas the Laplace overpressure for nanometric radii reaches tens to hundreds of MPa.
Under such conditions, an expansion written directly in excess pressures is usually not practically controlled.
By contrast, the liquid response may still satisfy the weak-compressibility condition \(|\beta_{T,l}\Delta P_l|\ll 1\) over a substantially broader interval.
This is why the planar-limit relation reported in the manuscript is taken from the formulation in relative density deviations, whereas the excess-pressure formulation remains an alternative local treatment of the same capillary--chemical balance whose practical usefulness depends on the regime.

Within the adopted derivation, vapor supersaturation is kept explicit until the planar limit is taken.
This point is physically important.
The part of the retained finite-curvature relation that does not contain the Tolman contribution determines the lowest-order supersaturation relation, whereas the Tolman-sensitive information enters only in the next curvature contribution kept in the derivation.
The planar-limit coefficient is obtained only after the planar limit is taken in that Tolman-sensitive relation.
Finite-curvature expressions that retain explicit supersaturation dependence are therefore useful for understanding the structure of the calculation, but they should not be identified with the strict planar-limit coefficient itself.
The water-specific analysis collected in Appendix~\ref{app:F} does not modify this derivation.
Its role is narrower: it supports \textit{a posteriori} that, for water in the regime emphasized here, the retained curvature contribution from which the Tolman term is extracted remains practically smaller than the explicitly retained lower-curvature part of the balance.

Against this background, earlier bulk-response approximations acquire a more precise interpretation.
Bartell's relation, Eq.~\eqref{eq:1.02}, follows from a pressure-linearized liquid description and therefore retains only the contribution associated with the liquid isothermal compressibility \(\beta_{T,l}\).
The Blokhuis--Kuipers expression, Eq.~\eqref{eq:1.03}, also goes beyond a purely liquid compressibility estimate because it includes both phases through the coexistence combination \(\rho_{0,l}^2\beta_{T,l}-\rho_{0,v}^2\beta_{T,v}\), but it still does not retain the explicit pressure derivative of the liquid compressibility.
In the adopted formulation, the planar-limit coefficient appears only after the linear two-phase coupling imposed by chemical equilibrium is combined with the second-order liquid relation between pressure and density.
If that additional liquid-state contribution is not retained, one is left with compressibility-only approximations and correspondingly shifted planar-limit estimates.
Our earlier result for interfaces closed in one or more intrinsic directions, Eq.~\eqref{eq:1.04}, is structurally closer to the present planar-limit relation because it already contains both \(\beta_{T,l}\) and \(\beta'_{T,l}\), although that earlier derivation is not itself a direct planar theory.
The comparison variants collected in Appendix~\ref{app:E} do not define alternative adopted derivations; their role is only algebraic and diagnostic, namely to clarify how different retained nonlinear contributions migrate into the retained \(h^2\)-sector and generate different planar coefficients.

The numerical consequences for water are substantial.
Applying the present treatment to water, using the IAPWS-IF97 industrial formulation together with homogeneous \((N,P,T)\) simulations of SPC/E and TIP4P/2005, shows that the planar-limit value of the Tolman length remains negative over the temperature range considered.
Its magnitude is of the order of \(-(0.7\text{--}1.3)\,\text{\AA}\), and its temperature dependence is markedly nonlinear, with a weakly nonmonotonic shape along the coexistence curve.
For the same thermodynamic input, the Bartell and Blokhuis--Kuipers approximations give substantially smaller magnitudes, \(|\delta_{\mathrm{Tolman}}^{\mathrm{planar}}|\simeq 0.30\text{--}0.36\,\text{\AA}\), together with nearly linear temperature trends.
This comparison shows that once the pressure dependence of the liquid compressibility is retained, both the magnitude and the temperature dependence of the planar-limit Tolman length change substantially relative to compressibility-only descriptions.

For SPC/E and TIP4P/2005 water at \(300\,\mathrm{K}\), the homogeneous \((N,P,T)\) fluctuation analysis yields mutually consistent negative planar-limit values of sub-\AA{} magnitude, close to \(-0.7\)~\AA.
An independent evaluation based on the IAPWS-IF97 industrial formulation gives \(\delta_{\mathrm{Tolman}}^{\mathrm{planar}}=-0.713\pm0.004\,\text{\AA}\) at the same coexistence state.
This agreement should be interpreted carefully.
It does not establish the absolute accuracy of any particular water model, nor does it remove the role of the external planar surface-tension input.
What it does show is that, once the pressure dependence of the liquid compressibility is retained, the planar-limit Tolman length can be obtained from bulk thermodynamic input and bulk volume statistics without resolving an explicit liquid--vapor interface.
At the same time, the block-analysis and shuffled diagnostics show that the fluctuation-based estimator is practically usable but statistically more demanding than the compressibility itself, because it depends explicitly on the third central moment \(m_3\).

Reported Tolman lengths for \textit{spherical} water droplets, obtained experimentally and in interface-resolved molecular simulations, are typically negative and of sub-\AA{} magnitude.
Azouzi \emph{et al.}~\cite{Azouzi2013} reported an experimental estimate of \(\delta_{\mathrm{Tolman}} \approx -0.47\,\text{\AA}\).
Molecular-dynamics studies using different water models reported \(\delta_{\mathrm{Tolman}}=-0.56 \pm 0.09\,\text{\AA}\)~\cite{Joswiak2013}, \(-0.56 \pm 0.10\,\text{\AA}\)~\cite{Joswiak2016}, \(-0.74\,\text{\AA}\)~\cite{Wilhelmsen2015}, and \(-0.50\,\text{\AA}\)~\cite{Kanduc2017}.
A recent simulation-based free-energy analysis for SPC, SPC/E, TIP4P, and TIP4P/2005 likewise found a consistently negative value, \(\delta_{\mathrm{Tolman}}\approx -0.48\,\text{\AA}\), and showed that accounting for curvature-dependent surface tension increases nucleation barriers relative to classical nucleation theory~\cite{doi:10.1021/jacs.5c14775}.
Although these studies concern curved droplets rather than the planar-limit quantity considered here, they provide context for the sign and order of magnitude of the present results.
The bulk formulation developed here therefore provides a complementary consistency check, without requiring explicit two-phase simulations.

The fluctuation representation makes the same point especially clear.
The planar-limit value of the Tolman length is not determined by the variance of the volume fluctuations alone.
In the \((N,P,T)\) ensemble, it depends on a specific combination of the second and third central moments of the volume distribution, and equivalently on the pressure response of the relative fluctuation width \(\sigma_V/\nu_1\).
The contribution associated with \(m_2/\nu_1\) corresponds to the compressibility contribution, whereas the contribution associated with \(m_3/m_2\) reflects the first correction associated with the asymmetry of the volume-fluctuation distribution.
In thermodynamic language, the same distinction appears as the difference between the compressibility itself and the pressure dependence of that compressibility.
The systematic difference between the full estimator and compressibility-only approximations is therefore not a minor numerical detail.
It reflects the fact that the first curvature correction is sensitive to the skewness-sensitive part of the bulk volume response.

In the mechanical description of a planar liquid--vapor interface, the planar-limit value of the Tolman length is the shift \(\delta_{\mathrm{Tolman}}^{\mathrm{planar}}=z_e-z_s\) between the equimolar surface and the surface of tension.
Here \(z_e\) is defined by vanishing adsorption excess, whereas \(z_s\) is fixed by the centered first-moment condition \(\int_{-\infty}^{\infty}(z-z_s)\,[P_N(z)-P_T(z)]\,\mathrm{d}z=0\).
Combined with the fluctuation representation, this means that the same planar-limit quantity has two complementary physical interpretations.
It is, on the one hand, a geometric characteristic of the interface and, on the other hand, a bulk linear-response quantity given by \((\partial/\partial P)_{T,N}\ln(\sigma_V/\nu_1)\big|_{P_0}\).

This distinction also clarifies the different physical roles of compressibility and Tolman-length response.
The isothermal compressibility, \(\beta_T\propto m_2/\nu_1\), is a purely bulk response determined by the second moment of the volume distribution.
In the thermodynamic limit, it does not by itself reveal either the presence or the geometry of an interface.
By contrast, the planar-limit value of the Tolman length carries explicit interfacial information.
Geometrically it is the shift between the surface of tension and the equimolar surface, while in the fluctuation formulation it is governed by the pressure response of the relative width of the volume distribution and therefore by a specific combination of the second and third moments of bulk volume fluctuations.
The fact that precisely this response reproduces the Tolman correction shows that the first curvature contribution is a bulk response that remains sensitive to the presence and structure of the interface.

The scope of the present treatment is therefore well defined.
It applies to one-component liquids in a neighborhood of liquid--vapor coexistence where the adopted local treatment remains controlled, that is, away from critical and spinodal regimes and provided that either an accurate thermal equation of state or sufficiently converged homogeneous \((N,P,T)\) volume statistics are available.
Systems in which the interfacial response is dominated by strong local ordering, adsorption, multicomponent composition gradients, or solid--liquid layering would require additional interface-specific information beyond the present one-component liquid--vapor formulation. To report the planar-limit Tolman length as a dimensional quantity, one additional input is required, namely an independently supplied planar surface tension.
Within these conditions, the present fluctuation-based bulk treatment is not restricted to water and can be transferred directly to other one-component liquids as an alternative to interface-resolved determinations of the first curvature correction to the thermodynamic surface tension.
\section{Conclusions}
\label{sec:conclusions}

We have shown that the planar-limit value of the Tolman length at liquid--vapor coexistence can be determined from bulk response properties of the homogeneous liquid.
The first curvature correction to the thermodynamic surface tension is therefore accessible without explicit resolution of a two-phase interface.

The essential physical point is that this planar-limit quantity is governed not only by the liquid compressibility itself, but also by the pressure dependence of that compressibility at coexistence.
In fluctuation terms, the relevant bulk information is carried not only by the width of the volume distribution in the isothermal--isobaric ensemble, but also by the first skewness-sensitive contribution.
The planar-limit Tolman length is therefore sensitive to the skewness-sensitive part of the bulk volume response.

We have also clarified the relation between the two local formulations of the same capillary--chemical balance considered in the manuscript.
The formulation written in excess pressures and the formulation written in relative density deviations are controlled by different smallness conditions and need not yield the same retained planar coefficient.
For weakly compressible water in the nanometric near-coexistence regime, the practically relevant formulation is the one based on relative density deviations, because the liquid response remains weak over a broader interval than the excess pressures themselves.

For water at \(300\,\mathrm{K}\), homogeneous \((N,P,T)\) simulations of SPC/E and TIP4P/2005 give mutually consistent negative planar-limit values close to \(-0.7\)~\AA.
An independent evaluation based on the IAPWS-IF97 industrial formulation gives \(-0.713\pm0.004\,\text{\AA}\) at the same coexistence state.
Along the coexistence curve, the same thermodynamic treatment predicts a negative planar-limit Tolman length with a weakly nonmonotonic temperature dependence.

The comparison with earlier compressibility-only approximations shows that the pressure dependence of the liquid compressibility makes a quantitatively essential contribution.
Omitting it leads to a systematic shift in both the magnitude and the temperature dependence of the predicted planar-limit Tolman length.

Taken together, these results establish that the planar-limit Tolman length remains an interfacial characteristic with a clear geometric meaning, but can nevertheless be inferred from bulk thermodynamic response and bulk volume statistics.
Because the value is inferred here from homogeneous bulk response rather than obtained from an explicitly resolved interface, the applicability of the framework is restricted to one-component liquid--vapor regimes where the local near-coexistence expansion remains controlled and where no additional interface-specific physics dominates the curvature response.
Within this regime, the present formulation provides a physically transparent bulk alternative to explicit interface-resolved determinations for one-component liquids near liquid--vapor coexistence.

\section*{Supplementary Material}
The supplementary material contains Supplementary Note~S1 with water-model parameters, Supplementary Note~S2 with detailed molecular-dynamics implementation settings and block-averaging diagnostics, and Supplementary Note~S3 with the IAPWS-IF97 and \(\gamma^\infty(T)\) tolerance model used for the coexistence-curve uncertainty propagation.

\section*{Acknowledgments}

Mykola Isaiev acknowledges support from the ANR project
``PROMENADE'' (Grant No.~ANR-23-CE50-0008).
Molecular simulations were performed using HPC resources from
GENCI-TGCC and GENCI-IDRIS through the eDARI Project Nos.~A0170913052
and A0190913052, as well as resources provided by the EXPLOR Center
hosted by the University of Lorraine.

\section*{Author Declarations}

\subsection*{Conflict of Interest}
The authors have no conflicts to disclose.

\subsection*{Ethics Approval}
This article does not contain any studies involving human participants or animals performed by any of the authors.

\subsection*{Author Contributions}
Sergii Burian: Conceptualization, Methodology, Software, Formal analysis, Investigation, Data curation, Visualization, Project administration, Writing -- original draft, Writing -- review \& editing.

Yevhenii Shportun: Software, Formal analysis, Validation, Visualization, Writing -- review \& editing.

Liudmyla Klochko: Methodology, Validation, Writing -- review \& editing.

Leonid Bulavin: Conceptualization, Supervision, Writing -- review \& editing.

Dmytro Gavryushenko: Methodology, Formal analysis, Validation, Writing -- review \& editing.

Mykola Isaiev: Conceptualization, Methodology, Formal analysis, Resources, Funding acquisition, Writing -- review \& editing.

\section*{Data Availability Statement}
The data and analysis scripts supporting the findings of this study are available from the corresponding author upon reasonable request. Additional simulation details, diagnostics, and uncertainty-propagation procedures are provided in the Supplementary Material.

\clearpage
\appendix

\renewcommand{\theequation}{\Alph{section}\arabic{equation}}
\makeatletter
\@addtoreset{equation}{section}
\@ifundefined{theHequation}{}{%
  \renewcommand{\theHsection}{app.\Alph{section}}%
  \renewcommand{\theHequation}{app.\Alph{section}.\arabic{equation}}%
}
\makeatother

\section{Volume fluctuations in the \((N,P,T)\) ensemble}
\label{app:A}

This Appendix collects the isothermal--isobaric fluctuation identities used in Sec.~\ref{sec:theory:fdt} to rewrite the planar-limit value of the same Tolman length in terms of moments of the volume distribution.
Its scope is limited to the bulk fluctuation representation: we specify the \((N,P,T)\) ensemble, derive the required pressure derivatives of the volume moments, and then recast the planar-limit relation into equivalent moment and fluctuation--response forms.

To characterize volume fluctuations in the \((N,P,T)\) ensemble, we consider the isothermal--isobaric partition function
\begin{equation}
\begin{split}
Q(T,P,N)
&=
\int_0^\infty \mathrm{d}V \sum_k
\exp\!\left[-\,\beta\bigl(PV + E_k(V;N)\bigr)\right].
\end{split}
\label{eq:SI-partition}
\end{equation}
Here \(E_k(V;N)\) is the energy of microstate \(k\) at volume \(V\) and particle number \(N\), and \(\beta = 1/(k_B T)\).

For observables depending only on \(V\), ensemble averages at fixed \((T,P,N)\) are given by
\begin{equation}
\begin{split}
\langle A \rangle
&=
\frac{1}{Q(T,P,N)}
\int_0^\infty \mathrm{d}V \sum_k A(V)
\\
&\quad \times
\exp\!\left[-\,\beta\bigl(PV + E_k(V;N)\bigr)\right].
\end{split}
\label{eq:SI-average}
\end{equation}
In particular, the mean volume is
\begin{equation}
\begin{split}
\nu_1(T,P,N)\equiv \langle V\rangle
&=
\frac{1}{Q(T,P,N)}
\int_0^\infty \mathrm{d}V \sum_k V
\\
&\quad \times
\exp\!\left[-\,\beta\bigl(PV + E_k(V;N)\bigr)\right].
\end{split}
\label{eq:SI-V-mean}
\end{equation}
Throughout this Appendix, angle brackets denote \((N,P,T)\)-ensemble averages.

\subsection{Pressure derivatives of raw and central moments}

It is convenient to begin with the pressure derivative of \(\ln Q\):
\begin{equation}
\begin{aligned}
\left(\frac{\partial \ln Q}{\partial P}\right)_{T,N}
&=
\frac{1}{Q}
\left(\frac{\partial Q}{\partial P}\right)_{T,N}
\\
&=
\frac{1}{Q}
\int_0^\infty \mathrm{d}V \sum_k
(-\beta V)\,
\exp\!\left[-\beta\bigl(PV+E_k(V;N)\bigr)\right]
\\
&=
-\beta \langle V\rangle
=
-\beta \nu_1.
\end{aligned}
\label{eq:SI-dlnQdP}
\end{equation}
Differentiating once more yields
\begin{equation}
\begin{split}
\left(\frac{\partial^2 \ln Q}{\partial P^2}\right)_{T,N}
&=
-\,\beta
\left(\frac{\partial \nu_1}{\partial P}\right)_{T,N}
\\
&=
\beta^2\Bigl[\langle V^2\rangle-(\langle V\rangle)^2\Bigr]
\\
&=
\beta^2 m_2,
\end{split}
\label{eq:SI-d2lnQdP2}
\end{equation}
where \(m_2\) is the second central moment, i.e., the variance of the volume distribution.

The second and third central moments used in the main text are
\begin{align}
m_2
&\equiv
\left\langle \bigl(V-\langle V\rangle\bigr)^2 \right\rangle
=
\langle V^2\rangle - \langle V\rangle^2,
\label{eq:SI-m2-def}
\\
m_3
&\equiv
\left\langle \bigl(V-\langle V\rangle\bigr)^3 \right\rangle
\nonumber\\
&=
\langle V^3\rangle
- 3\,\langle V\rangle\,\langle V^2\rangle
+ 2\,\langle V\rangle^3.
\label{eq:SI-m3-def}
\end{align}

\paragraph*{Derivative of the mean volume.}
Differentiating Eq.~\eqref{eq:SI-V-mean} with respect to \(P\), we obtain
\begin{align}
\left(\frac{\partial \nu_1}{\partial P}\right)_{T,N}
&=
\left(\frac{\partial \langle V\rangle}{\partial P}\right)_{T,N}
\nonumber\\
&=
\langle V(-\beta V)\rangle
- \langle V\rangle\,\langle -\beta V\rangle
\nonumber\\
&=
-\,\beta\Bigl[\langle V^2\rangle - \langle V\rangle^2\Bigr]
\nonumber\\
&=
-\,\beta m_2.
\label{eq:SI-dnu1dP}
\end{align}

\paragraph*{Derivative of the variance.}
Starting from \(m_2=\langle V^2\rangle-\nu_1^2\), we have
\begin{equation}
\left(\frac{\partial m_2}{\partial P}\right)_{T,N}
=
\left(\frac{\partial \langle V^2\rangle}{\partial P}\right)_{T,N}
- 2\nu_1
\left(\frac{\partial \nu_1}{\partial P}\right)_{T,N}.
\label{eq:SI-dm2dP-start}
\end{equation}
The derivative of the second raw moment is
\begin{align}
\left(\frac{\partial \langle V^2\rangle}{\partial P}\right)_{T,N}
&=
\langle V^2(-\beta V)\rangle
- \langle V^2\rangle\,\langle -\beta V\rangle
\nonumber\\
&=
-\,\beta\Bigl[\langle V^3\rangle - \langle V^2\rangle\,\langle V\rangle\Bigr].
\label{eq:SI-dV2dP}
\end{align}
Using Eq.~\eqref{eq:SI-dnu1dP} and substituting Eq.~\eqref{eq:SI-dV2dP} gives
\begin{align}
\left(\frac{\partial m_2}{\partial P}\right)_{T,N}
&=
-\,\beta\Bigl[\langle V^3\rangle - \langle V^2\rangle\,\langle V\rangle\Bigr]
- 2\nu_1(-\beta m_2)
\nonumber\\
&=
-\,\beta\Bigl[\langle V^3\rangle
- 3\,\langle V\rangle\,\langle V^2\rangle
+ 2\,\langle V\rangle^3\Bigr]
\nonumber\\
&=
-\,\beta m_3.
\label{eq:SI-dm2dP}
\end{align}
Equations~\eqref{eq:SI-dnu1dP} and~\eqref{eq:SI-dm2dP} can be summarized as
\begin{equation}
\left(\frac{\partial \nu_1}{\partial P}\right)_{T,N} = -\,\beta m_2,
\qquad
\left(\frac{\partial m_2}{\partial P}\right)_{T,N} = -\,\beta m_3.
\label{eq:SI-key-derivs}
\end{equation}

\subsection{Compressibility and its pressure derivative}

The isothermal compressibility is defined as
\begin{equation}
\beta_T(T,P)
=
-\,\frac{1}{\nu_1}
\left(\frac{\partial \nu_1}{\partial P}\right)_{T,N}.
\label{eq:SI-betaT-def}
\end{equation}
Using Eq.~\eqref{eq:SI-dnu1dP}, we obtain the standard fluctuation formula
\begin{equation}
\beta_T(T,P)
=
\frac{1}{k_B T}\,\frac{m_2}{\nu_1},
\label{eq:SI-betaT}
\end{equation}
which coincides with Eq.~\eqref{eq:2.16} in the main text.

Differentiating Eq.~\eqref{eq:SI-betaT} with respect to pressure gives
\begin{align}
\beta_T'(T,P)
&=
\left(\frac{\partial \beta_T}{\partial P}\right)_{T,N}
\nonumber\\
&=
\frac{1}{k_B T}
\left[
\frac{1}{\nu_1}
\left(\frac{\partial m_2}{\partial P}\right)_{T,N}
-
\frac{m_2}{\nu_1^2}
\left(\frac{\partial \nu_1}{\partial P}\right)_{T,N}
\right]
\nonumber\\
&=
\frac{1}{(k_B T)^2}
\left[
\left(\frac{m_2}{\nu_1}\right)^2
-
\frac{m_3}{\nu_1}
\right].
\label{eq:SI-betaT-prime}
\end{align}

\subsection{Fluctuation--response identity for the planar-limit relation}

Equations~\eqref{eq:SI-key-derivs} imply a compact identity for the pressure response of the relative width of the volume distribution.
Writing \(\sigma_V=\sqrt{m_2}\) and using \(\ln(\sigma_V/\nu_1)=\tfrac12\ln m_2-\ln\nu_1\), we obtain
\begin{align}
\left(\frac{\partial}{\partial P}\right)_{T,N}
\ln\!\left(\frac{\sigma_V}{\nu_1}\right)
&=
\frac{1}{2m_2}
\left(\frac{\partial m_2}{\partial P}\right)_{T,N}
-
\frac{1}{\nu_1}
\left(\frac{\partial \nu_1}{\partial P}\right)_{T,N}
\nonumber\\
&=
\beta\left(
\frac{m_2}{\nu_1}
-
\frac{m_3}{2m_2}
\right).
\label{eq:SI-fdt-sigma}
\end{align}
Equivalently,
\begin{equation}
\frac{1}{2k_B T}
\left(
2\,\frac{m_2}{\nu_1}
-
\frac{m_3}{m_2}
\right)
=
\left(\frac{\partial}{\partial P}\right)_{T,N}
\ln\!\left(\frac{\sigma_V}{\nu_1}\right),
\label{eq:SI-fdt-identity}
\end{equation}
where \(\sigma_V=\sqrt{m_2}\).

This fluctuation--response identity underlies the main-text expressions
\begin{equation}
\delta_{\mathrm{Tolman}}^{\mathrm{planar}}(T,P_0)
=
\left.
\frac{\gamma^\infty}{2k_B T}
\left(
2\,\frac{m_2}{\nu_1}
-
\frac{m_3}{m_2}
\right)
\right|_{P_0},
\label{eq:SI-delta-planar-moments}
\end{equation}
and
\begin{equation}
\frac{\delta_{\mathrm{Tolman}}^{\mathrm{planar}}}{\gamma^\infty}
=
\left.
\left(\frac{\partial}{\partial P}\right)_{T,N}
\ln\!\left(\frac{\sigma_V}{\nu_1}\right)
\right|_{P_0},
\label{eq:SI-delta-planar-fdt}
\end{equation}
which coincide with Eqs.~\eqref{eq:2.21} and~\eqref{eq:2.23} in the main text.

Here \(\delta_{\mathrm{Tolman}}^{\mathrm{planar}}\) denotes the planar-limit value of the same Tolman length, not a distinct interfacial quantity defined for a strictly planar interface.
All averages and derivatives appearing in Eqs.~\eqref{eq:SI-delta-planar-moments} and~\eqref{eq:SI-delta-planar-fdt} are evaluated at the coexistence state \((T,P_0,N)\), consistently with the curvature--compressibility analysis used in the main text.

\section{Local expansions of the chemical potential near planar coexistence}
\label{app:B}

This Appendix collects the local one-phase expansions of the chemical potential used as elementary building blocks in the later appendices.
Its role is preparatory: it establishes the local representations of \(\mu_i\) near the planar coexistence state in excess-pressure variables and in relative density deviations, without yet imposing capillary balance or two-phase closure.
In this sense, the Appendix does not constitute a separate derivation, but provides the local thermodynamic input that later enters both the pressure formulation and the adopted density-based formulation of the capillary--chemical balance.

\subsection{Coexistence reference state and local notation}
\label{app:B:setup}

Throughout this Appendix, temperature \(T\) is fixed and the reference state is the planar liquid--vapor coexistence point \((T,P_0)\), where
\begin{equation}
P_0=P_{\mathrm{sat}}(T),
\qquad
\mu_l(T,P_0)=\mu_v(T,P_0).
\label{eq:AB-coex}
\end{equation}
For each phase \(i\in\{l,v\}\), we write
\begin{equation}
\rho_{i0}\equiv \rho_i(T,P_0),
\qquad
\Delta P_i\equiv P_i-P_0,
\qquad
\Delta\rho_i\equiv \frac{\rho_i-\rho_{i0}}{\rho_{i0}}.
\label{eq:AB-defs}
\end{equation}
Unless stated otherwise, all pressure derivatives in this Appendix are taken at fixed temperature and evaluated at the coexistence reference state \((T,P_0)\).

We also use
\begin{equation}
\beta_{T,i}\equiv \beta_{T,i}(T,P_0),
\qquad
\beta'_{T,i}\equiv
\left(\frac{\partial \beta_{T,i}}{\partial P}\right)_T\Bigg|_{P_0},
\label{eq:AB-beta}
\end{equation}
and, when needed later,
\begin{equation}
A_i \equiv 1+\frac{\beta'_{T,i}}{\beta_{T,i}^2}.
\label{eq:AB-Ai}
\end{equation}
The purpose of the present Appendix is only to establish local one-phase expansions of \(\mu_i\) in the pressure and density variables defined above.
No coupling between the liquid and vapor branches is imposed here.

\subsection{Pressure derivatives of the chemical potential}
\label{app:B:muPderivs}

Along an isotherm, the chemical potential satisfies the standard relation
\begin{equation}
\left(\frac{\partial \mu_i}{\partial P}\right)_T=\frac{1}{\rho_i}.
\label{eq:AB-muP1}
\end{equation}
Using
\begin{equation}
\left(\frac{\partial \rho_i}{\partial P}\right)_T
=
\beta_{T,i}\rho_i,
\label{eq:AB-rhoP}
\end{equation}
the second pressure derivative becomes
\begin{equation}
\left(\frac{\partial^2 \mu_i}{\partial P^2}\right)_T
=
-\frac{\beta_{T,i}}{\rho_i}.
\label{eq:AB-muP2}
\end{equation}
Differentiating once more gives
\begin{equation}
\left(\frac{\partial^3 \mu_i}{\partial P^3}\right)_T
=
\frac{\beta_{T,i}^2-\beta'_{T,i}}{\rho_i}.
\label{eq:AB-muP3}
\end{equation}
Evaluated at the coexistence reference state, these relations read
\begin{equation}
\begin{aligned}
\mu'_{i,0} &= \frac{1}{\rho_{i0}},
\\
\mu''_{i,0} &= -\frac{\beta_{T,i}}{\rho_{i0}},
\\
\mu'''_{i,0} &=
\frac{\beta_{T,i}^{2}-\beta'_{T,i}}{\rho_{i0}},
\end{aligned}
\label{eq:AB-muP-refs}
\end{equation}
where, for compactness, the subscript \(0\) indicates evaluation at \((T,P_0)\).

These pressure derivatives provide the local expansion coefficients used later when the chemical-potential equality is expanded directly in excess-pressure variables.

\subsection{Local pressure expansion of the chemical potential}
\label{app:B:muPseries}

Expanding \(\mu_i(T,P)\) about \(P=P_0\) gives
\begin{equation}
\begin{split}
\mu_i(P_0+\Delta P_i)
&=
\mu_i(P_0)
+
\mu'_{i,0}\Delta P_i
+
\frac12\,\mu''_{i,0}\Delta P_i^2
\\
&\quad
+
\frac16\,\mu'''_{i,0}\Delta P_i^3
+
O(\Delta P_i^4).
\end{split}
\label{eq:AB-muP-Taylor}
\end{equation}
Using Eq.~\eqref{eq:AB-muP-refs}, this becomes
\begin{equation}
\begin{split}
\mu_i(P_0+\Delta P_i)
&=
\mu_i(P_0)
+
\frac{\Delta P_i}{\rho_{i0}}
-
\frac{\beta_{T,i}}{2\rho_{i0}}\,\Delta P_i^2
\\
&\quad
+
\frac{\beta_{T,i}^2-\beta'_{T,i}}{6\rho_{i0}}\,\Delta P_i^3
+
O(\Delta P_i^4).
\end{split}
\label{eq:AB-muP-series}
\end{equation}

Equation~\eqref{eq:AB-muP-series} is the local pressure expansion used later when the chemical-potential equality is expanded in excess-pressure variables.
In particular, the liquid and vapor branches may then be evaluated at different shifted pressure arguments, but the one-phase expansion itself is the local result recorded here.

\subsection{Density derivatives of the chemical potential}
\label{app:B:muRderivs}

We next construct the local density representation of \(\mu_i\).
Using the chain rule together with Eqs.~\eqref{eq:AB-muP1} and~\eqref{eq:AB-rhoP}, we obtain
\begin{equation}
\left(\frac{\partial \mu_i}{\partial \rho_i}\right)_T
=
\frac{
\left(\frac{\partial \mu_i}{\partial P}\right)_T
}{
\left(\frac{\partial \rho_i}{\partial P}\right)_T
}
=
\frac{1}{\beta_{T,i}\rho_i^2}.
\label{eq:AB-muR1}
\end{equation}
Differentiating with respect to \(\rho_i\) at fixed \(T\) gives
\begin{equation}
\left(\frac{\partial^2 \mu_i}{\partial \rho_i^2}\right)_T
=
-
\frac{2+\beta'_{T,i}/\beta_{T,i}^2}{\beta_{T,i}\rho_i^3}.
\label{eq:AB-muR2}
\end{equation}
Equivalently, using Eq.~\eqref{eq:AB-Ai},
\begin{equation}
\left(\frac{\partial^2 \mu_i}{\partial \rho_i^2}\right)_T
=
-
\frac{1+A_i}{\beta_{T,i}\rho_i^3}.
\label{eq:AB-muR2-A}
\end{equation}
Evaluated at the coexistence reference state, the corresponding coefficients are
\begin{equation}
\begin{aligned}
\left(\frac{\partial \mu_i}{\partial \rho_i}\right)_T\Bigg|_{\rho_{i0}}
&=
\frac{1}{\beta_{T,i}\rho_{i0}^2},
\\
\left(\frac{\partial^2 \mu_i}{\partial \rho_i^2}\right)_T\Bigg|_{\rho_{i0}}
&=
-
\frac{1+A_i}{\beta_{T,i}\rho_{i0}^3}.
\end{aligned}
\label{eq:AB-muR-refs}
\end{equation}

These relations provide the local density derivatives required for the later expansion of the chemical-potential equality in relative density deviations.

\subsection{Local density expansion of the chemical potential}
\label{app:B:muRseries}

Writing
\begin{equation}
\rho_i=\rho_{i0}(1+\Delta\rho_i),
\label{eq:AB-rho-rel}
\end{equation}
we expand \(\mu_i(\rho_i;T)\) about \(\rho_{i0}\):
\begin{equation}
\begin{split}
\mu_i\!\bigl(\rho_{i0}(1+\Delta\rho_i)\bigr)
&=
\mu_i(\rho_{i0})
+
\left.\left(\frac{\partial \mu_i}{\partial \rho_i}\right)_T\right|_{\rho_{i0}}
\,\rho_{i0}\Delta\rho_i
\\
&\quad
+
\frac12
\left.\left(\frac{\partial^2 \mu_i}{\partial \rho_i^2}\right)_T\right|_{\rho_{i0}}
\,\rho_{i0}^2\Delta\rho_i^2
\\
&\quad
+
O(\Delta\rho_i^3).
\end{split}
\label{eq:AB-muR-Taylor}
\end{equation}
Using Eq.~\eqref{eq:AB-muR-refs}, we obtain
\begin{equation}
\begin{aligned}
\mu_i\!\bigl(\rho_{i0}(1+\Delta\rho_i)\bigr)
&=
\mu_i(\rho_{i0})
+
\frac{\Delta\rho_i}{\beta_{T,i}\rho_{i0}}
\\
&\quad
-
\frac{1}{2\beta_{T,i}\rho_{i0}}
\left(
2+\frac{\beta'_{T,i}}{\beta_{T,i}^2}
\right)\Delta\rho_i^2
+
O(\Delta\rho_i^3)
\\
&=
\mu_i(\rho_{i0})
+
\frac{\Delta\rho_i}{\beta_{T,i}\rho_{i0}}
-
\frac{1+A_i}{2\beta_{T,i}\rho_{i0}}\Delta\rho_i^2
\\
&\quad
+
O(\Delta\rho_i^3).
\end{aligned}
\label{eq:AB-muR-series}
\end{equation}

Equation~\eqref{eq:AB-muR-series} is the local density expansion used later when the chemical-potential equality is expanded in relative density deviations.
At this stage, however, it remains entirely one-phase and does not yet impose any relation between the liquid and vapor branches.

\subsection{Illustrative coexistence and shifted-pressure chemical-potential branches}
\label{app:B:illustration}

\begin{figure*}[t]
	\centering
	\includegraphicsmaybe[width=\textwidth]{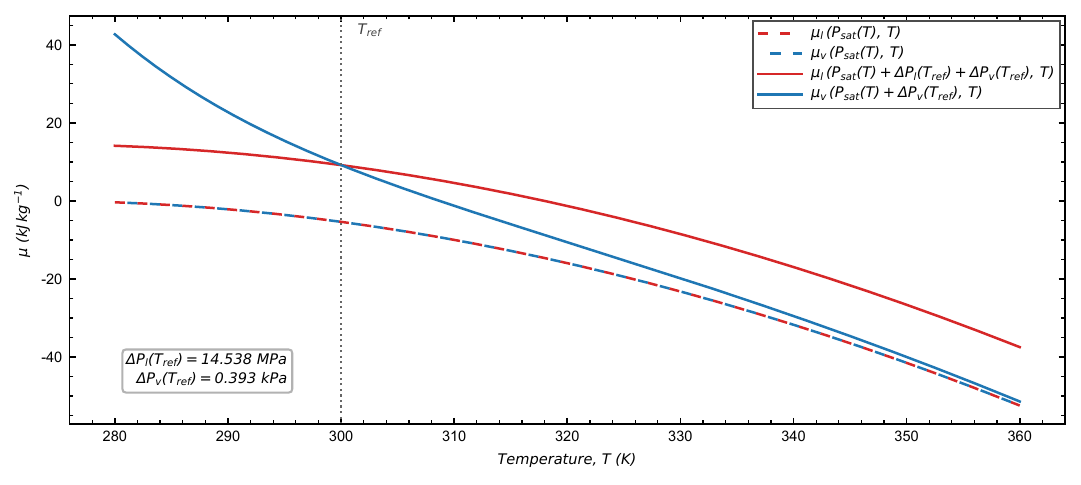}
	\caption{\label{fig:mu-capillary-4curves}
	Chemical potentials \(\mu_l\) and \(\mu_v\) of water as functions of temperature \(T\), evaluated from the IAPWS-IF97 industrial formulation.
	Dashed lines show \(\mu_l(P_{\mathrm{sat}}(T),T)\) and \(\mu_v(P_{\mathrm{sat}}(T),T)\) along planar liquid--vapor coexistence, where \(P_0=P_{\mathrm{sat}}(T)\) and \(\mu_l=\mu_v\).
	Solid lines show the same quantities evaluated at the shifted pressures \(P_{\mathrm{sat}}(T)+\Delta P_l(T_{\mathrm{ref}})\) and \(P_{\mathrm{sat}}(T)+\Delta P_v(T_{\mathrm{ref}})\) corresponding to a spherical droplet with \(R=10\,\mathrm{nm}\), \(\delta=-0.7\,\text{\AA}\), and \(T_{\mathrm{ref}}=300\,\mathrm{K}\), for which \(\Delta P_l(T_{\mathrm{ref}})=14.538\,\mathrm{MPa}\) and \(\Delta P_v(T_{\mathrm{ref}})=0.393\,\mathrm{kPa}\).
	The construction illustrates how curvature-induced pressure shifts displace the liquid and vapor branches away from their planar coexistence values; the shifts are held fixed at their reference-state values and the figure is intended only as a geometric illustration.}
\end{figure*}

Although the present Appendix is devoted to local one-phase expansions, it is useful to visualize the chemical-potential branches that later enter the capillary--chemical balance through shifted pressure arguments.
Figure~\ref{fig:mu-capillary-4curves} contrasts the planar coexistence branches with the branches evaluated at the shifted liquid and vapor pressures corresponding to a representative curved droplet.

This figure is illustrative only.
It is not part of the derivation of the local one-phase expansions recorded above.
Its role is to clarify how the shifted liquid and vapor arguments used later in the pressure-side and density-based expansions are related to the planar coexistence reference state.

\subsection{Role in the later appendices}
\label{app:B:role}

The results of the present Appendix serve only as local one-phase input for the later derivations.
Equation~\eqref{eq:AB-muP-series} supplies the one-phase pressure expansion of the chemical potential used in Appendix~\ref{app:D}, where the same capillary--chemical balance is expanded in excess-pressure variables.
Equation~\eqref{eq:AB-muR-series} supplies the one-phase density expansion of the chemical potential used in Appendix~\ref{app:E}, where the same equilibrium conditions are expanded in relative density deviations.

Thus, the present Appendix should be read only as a preparatory local block.
It does not yet impose two-phase chemical coupling, capillary balance, or any retained mixed finite-curvature relation.
Those additional steps enter only in the later appendices, where the local one-phase expansions recorded here are combined into the specific formulations used in the manuscript.

\section{One-phase excess-pressure / relative-density series and local inversion}
\label{app:C}

This Appendix derives the local one-phase relations between excess pressure and relative density deviation along an isotherm near planar coexistence.
Its role is again preparatory, but now at the level of the thermal equation of state: first the forward series \(\Delta P_i(\Delta\rho_i)\) is constructed, and then the corresponding local inverse series \(\Delta\rho_i(\Delta P_i)\) is obtained.
These expansions are later used in the adopted density-based derivation, while the water-based plots included here support that the retained truncation orders remain adequate over the regime emphasized in the manuscript.

\subsection{Reference state and one-phase variables}
\label{app:C:setup}

Throughout this Appendix, temperature \(T\) is fixed and the reference state is the planar liquid--vapor coexistence point \((T,P_0)\), where
\begin{equation}
P_0=P_{\mathrm{sat}}(T),
\qquad
\mu_l(T,P_0)=\mu_v(T,P_0).
\label{eq:AC-coex}
\end{equation}
For each phase \(i\in\{l,v\}\), we define
\begin{equation}
\rho_{i0}\equiv \rho_i(T,P_0),
\qquad
\Delta P_i\equiv P_i-P_0,
\qquad
\Delta\rho_i\equiv \frac{\rho_i-\rho_{i0}}{\rho_{i0}}.
\label{eq:AC-defs}
\end{equation}
We also write
\begin{equation}
\begin{aligned}
\beta_i &\equiv \beta_{T,i}(T,P_0),
\\
\beta_i' &\equiv \left(\frac{\partial \beta_{T,i}}{\partial P}\right)_T\Bigg|_{P_0},
\\
\beta_i'' &\equiv \left(\frac{\partial^2 \beta_{T,i}}{\partial P^2}\right)_T\Bigg|_{P_0}.
\end{aligned}
\label{eq:AC-beta}
\end{equation}
For later convenience, we introduce the compact combination
\begin{equation}
A_i \equiv 1+\frac{\beta_i'}{\beta_i^2}.
\label{eq:AC-Ai}
\end{equation}

At the present stage, all relations are purely one-phase and local.
No two-phase chemical coupling, capillary balance, or finite-curvature mixed relation is imposed in this Appendix.

\subsection{Forward one-phase series: \texorpdfstring{$\Delta P_i(\Delta\rho_i)$}{DeltaPi(DeltaRhoi)}}
\label{app:C:forward}

We begin from the local thermal equation of state of phase \(i\), written along an isotherm as \(P_i=P_i(\rho;T)\).
The one-phase excess pressure is then
\begin{equation}
\Delta P_i
=
P_i\!\bigl(\rho_{i0}(1+\Delta\rho_i);T\bigr)-P_i(\rho_{i0};T).
\label{eq:AC-DeltaPi-def}
\end{equation}

Along an isotherm, the standard identities
\begin{equation}
\left(\frac{\partial \mu_i}{\partial P}\right)_T=\frac{1}{\rho_i},
\qquad
\left(\frac{\partial \rho_i}{\partial P}\right)_T=\beta_{T,i}(T,P)\rho_i
\label{eq:AC-iso-identities}
\end{equation}
imply
\begin{equation}
\left(\frac{\partial P}{\partial \rho_i}\right)_T
=
\frac{1}{\beta_{T,i}(T,P)\rho_i}.
\label{eq:AC-dPdrho}
\end{equation}
Evaluating this derivative at the reference state gives
\begin{equation}
P_i'(\rho_{i0})
=
\frac{1}{\beta_i\rho_{i0}}.
\label{eq:AC-Pprime}
\end{equation}

Differentiating Eq.~\eqref{eq:AC-dPdrho} once more with respect to \(\rho_i\) at fixed \(T\), and then evaluating at \(\rho_{i0}\), yields
\begin{equation}
P_i''(\rho_{i0})
=
-\,\frac{1+\beta_i'/\beta_i^2}{\beta_i\rho_{i0}^2}
=
-\,\frac{A_i}{\beta_i\rho_{i0}^2}.
\label{eq:AC-Pdoubleprime}
\end{equation}
A further differentiation gives
\begin{equation}
P_i^{(3)}(\rho_{i0})
=
\frac{1}{\beta_i\rho_{i0}^3}
\left[
2+\frac{3\beta_i'}{\beta_i^2}
+\frac{3(\beta_i')^2}{\beta_i^4}
-\frac{\beta_i''}{\beta_i^3}
\right].
\label{eq:AC-P3}
\end{equation}

Expanding \(P_i(\rho;T)\) about \(\rho_{i0}\) then gives
\begin{equation}
\begin{split}
\Delta P_i
&=
P_i'(\rho_{i0})\,\rho_{i0}\Delta\rho_i
+
\frac12\,P_i''(\rho_{i0})\,\rho_{i0}^2\Delta\rho_i^2
\\
&\quad
+
\frac16\,P_i^{(3)}(\rho_{i0})\,\rho_{i0}^3\Delta\rho_i^3
+
O(\Delta\rho_i^4).
\end{split}
\label{eq:AC-DeltaPi-Taylor}
\end{equation}
Using Eqs.~\eqref{eq:AC-Pprime}--\eqref{eq:AC-P3}, we obtain
\begin{equation}
\begin{aligned}
\Delta P_i
&=
\frac{\Delta\rho_i}{\beta_i}
-
\frac{1}{2\beta_i}
\left(
1+\frac{\beta_i'}{\beta_i^2}
\right)\Delta\rho_i^2
\\
&\quad
+
\frac{1}{6\beta_i}
\left[
2+\frac{3\beta_i'}{\beta_i^2}
+\frac{3(\beta_i')^2}{\beta_i^4}
-\frac{\beta_i''}{\beta_i^3}
\right]\Delta\rho_i^3
+
O(\Delta\rho_i^4)
\\
&=
\frac{\Delta\rho_i}{\beta_i}
-
\frac{A_i}{2\beta_i}\Delta\rho_i^2
\\
&\quad
+
\frac{1}{6\beta_i}
\left[
2+\frac{3\beta_i'}{\beta_i^2}
+\frac{3(\beta_i')^2}{\beta_i^4}
-\frac{\beta_i''}{\beta_i^3}
\right]\Delta\rho_i^3
+
O(\Delta\rho_i^4).
\end{aligned}
\label{eq:AC-DeltaPi-series}
\end{equation}

Equation~\eqref{eq:AC-DeltaPi-series} is the local forward one-phase series used later as the basic thermal-equation-of-state input for the adopted density-based formulation.
At this stage, however, it remains entirely one-phase and does not yet encode any coupling between the liquid and vapor branches.

\subsection{Local inverse series: \texorpdfstring{$\Delta\rho_i(\Delta P_i)$}{DeltaRhoi(DeltaPi)}}
\label{app:C:inverse}

We next invert Eq.~\eqref{eq:AC-DeltaPi-series} to the order required later in the manuscript.
We seek a local inverse relation of the form
\begin{equation}
\Delta\rho_i
=
a_1\Delta P_i+a_2\Delta P_i^2+O(\Delta P_i^3).
\label{eq:AC-inv-ansatz}
\end{equation}
Substituting Eq.~\eqref{eq:AC-inv-ansatz} into Eq.~\eqref{eq:AC-DeltaPi-series} and matching equal powers of \(\Delta P_i\), we obtain
\begin{equation}
a_1=\beta_i,
\qquad
a_2=\frac12
\left(
1+\frac{\beta_i'}{\beta_i^2}
\right)\beta_i^2
=
\frac12\,A_i\beta_i^2.
\label{eq:AC-inv-coeffs}
\end{equation}
Therefore,
\begin{equation}
\Delta\rho_i
=
\beta_i\Delta P_i
+
\frac12
\left(
1+\frac{\beta_i'}{\beta_i^2}
\right)
(\beta_i\Delta P_i)^2
+
O(\Delta P_i^3).
\label{eq:AC-Deltarho-of-DeltaP}
\end{equation}

For compactness, it is useful to introduce the one-phase dimensionless excess-pressure variable
\begin{equation}
\pi_i\equiv \beta_i\Delta P_i.
\label{eq:AC-pii}
\end{equation}
Then Eq.~\eqref{eq:AC-Deltarho-of-DeltaP} may be written as
\begin{equation}
\begin{aligned}
\Delta\rho_i
&=
\pi_i
+
\frac12
\left(
1+\frac{\beta_i'}{\beta_i^2}
\right)\pi_i^2
+
O(\pi_i^3)
\\
&=
\pi_i+\frac12\,A_i\pi_i^2+O(\pi_i^3).
\end{aligned}
\label{eq:AC-Deltarho-pi}
\end{equation}

Equation~\eqref{eq:AC-Deltarho-pi} is the local inverse one-phase series used later in the adopted density-based derivation.
In the adopted manuscript-level reduction, the liquid branch is retained to second order, whereas the vapor branch is later kept only at leading order.
That asymmetry, however, is not imposed here; the present Appendix records only the local one-phase inversion itself.

\subsection{Water-based support for the retained local truncation orders}
\label{app:C:water}

The one-phase local series derived above are used later only at low retained orders.
For the water regime emphasized in the manuscript, it is therefore useful to support that these retained orders remain adequate over the interval of practical interest.
The present subsection serves exactly that purpose.

\begin{figure*}[t]
\centering
\includegraphicsmaybe[width=\textwidth]{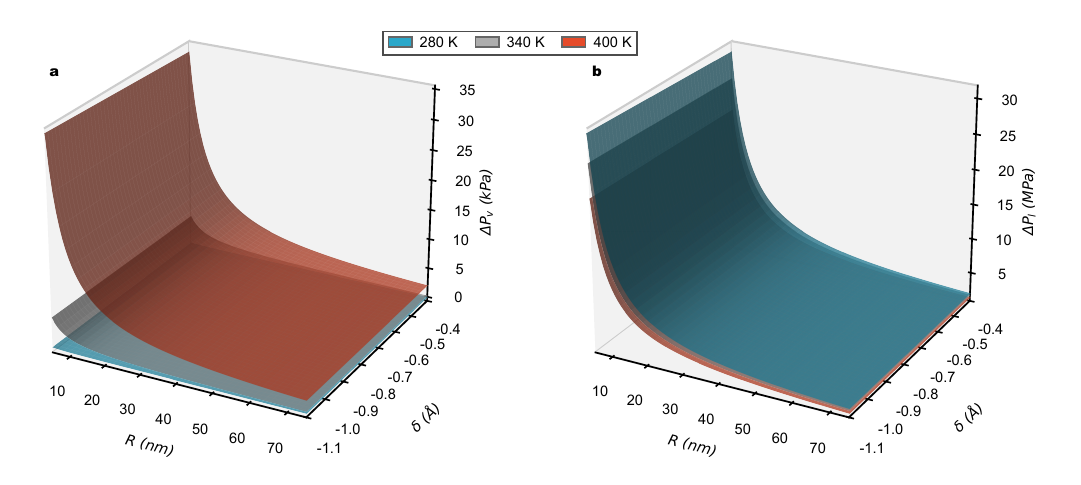}
\caption{\label{fig:AC-excess-pressures}
Excess vapor and liquid pressures for spherical droplets.
Excess vapor and liquid pressures \(\Delta P_v\) and \(\Delta P_l\) for spherical droplets as functions of droplet radius \(R\), assumed Tolman length \(\delta\), and temperature \(T\), evaluated from the IAPWS-IF97 industrial formulation.
Panel (a) shows \(\Delta P_v\) in kPa and panel (b) shows \(\Delta P_l\) in MPa over \(R=5\text{--}75\,\mathrm{nm}\), \(\delta=-1.1\text{ to }-0.4\,\text{\AA}\), and \(T=280,340,400\,\mathrm{K}\).}
\end{figure*}

\begin{figure*}[t]
\centering
\includegraphicsmaybe[width=\textwidth]{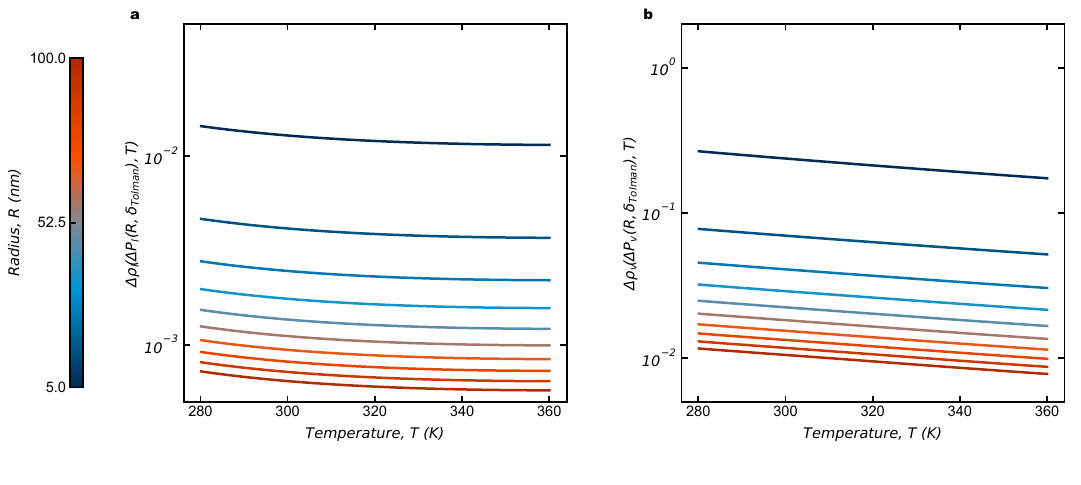}
\caption{\label{fig:AC-rel-rho-excess}
Relative density deviations in liquid and vapor.
Relative density deviations \(\Delta\rho_l\) and \(\Delta\rho_v\), defined with respect to planar liquid--vapor coexistence \((T,P_0)\), are shown for water as functions of temperature \(T\).
The values were evaluated from the IAPWS-IF97 industrial formulation using capillary-equilibrium pressure shifts for spherical droplets with \(\delta_{\mathrm{Tolman}}=-0.7\,\text{\AA}\).
Panel (a) shows the liquid response, and panel (b) shows the vapor response.
Colors indicate the droplet radius \(R\), and both panels use logarithmic vertical scales.}
\end{figure*}

Figure~\ref{fig:AC-excess-pressures} shows the scale of the one-phase excess-pressure variables for water, using thermodynamic properties evaluated from the IAPWS-IF97 industrial formulation as representative empirical input.
Figure~\ref{fig:AC-rel-rho-excess} shows the corresponding scale of the one-phase relative density deviations.
Taken together, these plots support the manuscript-level strategy: even when excess pressures are not small on the coexistence-pressure scale, the retained one-phase relative-density deviations may remain sufficiently small for the adopted local density-side truncation to stay meaningful over the water regime emphasized here.

These figures are therefore not merely illustrative.
Their role is to support, on the water example, that the particular truncation levels later retained in the one-phase liquid and vapor expansions are adequate for the regime to which the manuscript-level derivation is applied.

\subsection{Use in the later derivations}
\label{app:C:role}

The results of the present Appendix are used later only as local one-phase input.
Equation~\eqref{eq:AC-DeltaPi-series} provides the forward one-phase thermal-equation-of-state series, whereas Eqs.~\eqref{eq:AC-Deltarho-of-DeltaP} and~\eqref{eq:AC-Deltarho-pi} provide the corresponding local inverse series.

In the later density-based derivation, these inverse one-phase relations are combined with the leading two-phase chemical coupling and only afterward with capillary balance.
Thus, the present Appendix does not yet construct any two-phase balance or any finite-curvature relation.
It supplies only the local one-phase pressure--density relations that are later inserted into the adopted density-based derivation.
\section{Pressure-side expansion of the capillary--chemical balance}
\label{app:D}

This Appendix expands the pressure-side formulation discussed in Sec.~\ref{sec:theory:pressure}.
Starting from the same capillary--chemical balance as in the main text, it writes the local expansion directly in excess-pressure variables about the planar coexistence state and makes explicit the role of the shifted liquid argument \(P_0+\Delta P_v+\Delta P\).
Its purpose is to document the pressure-side supersaturation relation, the planar-limit coefficient obtained within the minimal pressure-side formulation, and the corresponding finite-curvature continuations obtained when explicit supersaturation dependence is retained.

\subsection{Pressure-side setup and shifted arguments}
\label{app:D:setup}

Throughout this Appendix, temperature \(T\) is fixed and the reference state is the planar coexistence point \((T,P_0)\), where
\begin{equation}
P_0=P_{\mathrm{sat}}(T),
\qquad
\mu_l(T,P_0)=\mu_v(T,P_0).
\label{eq:AD-coex}
\end{equation}
We define the phase excess pressures and the capillary pressure difference by
\begin{equation}
\Delta P_v\equiv P_v-P_0,
\qquad
\Delta P_l\equiv P_l-P_0,
\qquad
\Delta P\equiv P_l-P_v,
\label{eq:AD-defs}
\end{equation}
so that
\begin{equation}
\Delta P_l=\Delta P_v+\Delta P.
\label{eq:AD-DeltaPl}
\end{equation}
With the sign convention adopted in Sec.~\ref{sec:theory:setup}, this corresponds to a liquid droplet or liquid filament surrounded by vapor, so that the liquid pressure exceeds the vapor pressure by the Laplace amount.

The pressure-side formulation uses the same chemical-equilibrium condition as in the main text,
\begin{equation}
\mu_v(P_0+\Delta P_v)=\mu_l(P_0+\Delta P_l),
\label{eq:AD-chem}
\end{equation}
but the key operational point is how the two branches are expanded.
On the vapor side, the local argument is \(P_0+\Delta P_v\).
On the liquid side, the local argument is \(P_0+\Delta P_l=P_0+\Delta P_v+\Delta P\).
Thus, the liquid branch is not expanded in the capillary increment alone, but at the shifted liquid state determined jointly by vapor supersaturation and capillary overpressure.

All response coefficients below are evaluated at the coexistence reference state \((T,P_0)\), and we write
\begin{equation}
\begin{aligned}
\beta_{T,i} &\equiv \beta_{T,i}(T,P_0),
\\
\beta'_{T,i} &\equiv
\left(\frac{\partial \beta_{T,i}}{\partial P}\right)_T\Bigg|_{P_0},
\\
\beta''_{T,i} &\equiv
\left(\frac{\partial^2 \beta_{T,i}}{\partial P^2}\right)_T\Bigg|_{P_0}.
\end{aligned}
\label{eq:AD-beta}
\end{equation}
For compactness, we also introduce
\begin{equation}
\begin{aligned}
r_\rho &\equiv \frac{\rho_{v0}}{\rho_{l0}},
&
h &\equiv H\gamma^\infty\beta_{T,l},
\\
\Pi_v &\equiv \beta_{T,l}\Delta P_v,
&
\Delta &\equiv \frac{\delta_{\mathrm{Tolman}}}{\gamma^\infty\beta_{T,l}},
\end{aligned}
\label{eq:AD-dimless}
\end{equation}
together with
\begin{equation}
A_l\equiv 1+\frac{\beta'_{T,l}}{\beta_{T,l}^2}.
\label{eq:AD-Al}
\end{equation}

\subsection{Local pressure-expanded capillary--chemical balance}
\label{app:D:local}

At fixed \(T\), the pressure derivatives of the chemical potential satisfy
\begin{equation}
\begin{aligned}
\left(\frac{\partial \mu_i}{\partial P}\right)_T
&=
\frac{1}{\rho_i},
\\[2pt]
\left(\frac{\partial^2 \mu_i}{\partial P^2}\right)_T
&=
-\frac{\beta_{T,i}}{\rho_i},
\\[2pt]
\left(\frac{\partial^3 \mu_i}{\partial P^3}\right)_T
&=
\frac{\beta_{T,i}^2-\beta'_{T,i}}{\rho_i}.
\end{aligned}
\label{eq:AD-mu-derivs}
\end{equation}
Hence, for either phase \(i\in\{l,v\}\), the local pressure expansion about \(P_0\) is
\begin{equation}
\begin{split}
\mu_i(P_0+\Delta P_i)
&=
\mu_i(P_0)
+
\frac{\Delta P_i}{\rho_{i0}}
-
\frac{\beta_{T,i}}{2\rho_{i0}}\Delta P_i^2
\\
&\quad
+
\frac{\beta_{T,i}^2-\beta'_{T,i}}{6\rho_{i0}}\Delta P_i^3
+
O(\Delta P_i^4),
\end{split}
\label{eq:AD-mui-series}
\end{equation}
where \(\rho_{i0}\equiv \rho_i(T,P_0)\).

In the present pressure-side formulation, Eq.~\eqref{eq:AD-mui-series} is applied to the vapor branch with \(\Delta P_i=\Delta P_v\) and to the liquid branch with \(\Delta P_i=\Delta P_l=\Delta P_v+\Delta P\).
Substituting Eq.~\eqref{eq:AD-mui-series} into Eq.~\eqref{eq:AD-chem}, we obtain the local pressure-expanded capillary--chemical balance
\begin{equation}
\begin{split}
0
&=
\frac{\Delta P_l}{\rho_{l0}}
-
\frac{\Delta P_v}{\rho_{v0}}
-
\frac{\beta_{T,l}}{2\rho_{l0}}\,\Delta P_l^2
+
\frac{\beta_{T,v}}{2\rho_{v0}}\,\Delta P_v^2
\\
&\quad
+
\frac{\beta_{T,l}^2-\beta'_{T,l}}{6\rho_{l0}}\,\Delta P_l^3
-
\frac{\beta_{T,v}^2-\beta'_{T,v}}{6\rho_{v0}}\,\Delta P_v^3
\\
&\quad
+
O(\Delta P_l^4,\Delta P_v^4).
\end{split}
\label{eq:AD-local-balance}
\end{equation}

Equation~\eqref{eq:AD-local-balance} is the pressure-side starting point of this Appendix.
Once the capillary relation is imposed, different retained pressure-side formulations correspond to keeping different subsets of this local relation and then reorganizing the result in the mixed variables \((h,\Pi_v,\Delta)\).

\subsection{Minimal pressure-side formulation}
\label{app:D:minimal}

We now combine the local pressure-expanded balance with the Tolman-corrected capillary relation
\begin{equation}
\Delta P
=
H\gamma^\infty\bigl(1-H\delta_{\mathrm{Tolman}}\bigr).
\label{eq:AD-cap}
\end{equation}

A point that matters for interpretation is that, in the actual symbolic workflow, the first retained mixed pressure-side relation already contains explicit \(\Pi_v\)-dependence, because the liquid branch is evaluated at the shifted argument \(P_0+\Delta P_v+\Delta P\).
Accordingly, the planar-limit coefficient is not obtained from a prior separate pressure-only symbolic balance.
Instead, it is obtained within a minimal pressure-side formulation of the retained mixed structure: one keeps the leading supersaturation sector together with the \(\Pi_v\)-independent part of the retained \(h^2\)-sector.

This gives
\begin{equation}
\left(\frac12+\Delta\right)h^2
-
h
-
\Pi_v
+
\frac{\Pi_v}{r_\rho}
=
0.
\label{eq:AD-min-balance}
\end{equation}

Equation~\eqref{eq:AD-min-balance} yields two distinct results.
First, the leading terms give the supersaturation relation
\begin{equation}
\Pi_v
=
\frac{h\,r_\rho}{1-r_\rho}
+
O(h^2),
\label{eq:AD-Piv-min}
\end{equation}
that is,
\begin{equation}
\Delta P_v
=
H\gamma^\infty\,
\frac{\rho_{v0}}{\rho_{l0}-\rho_{v0}}
+
O(H^2).
\label{eq:AD-DeltaPv-leading}
\end{equation}
This is the pressure-side form of the leading Kelvin-type supersaturation scaling quoted in the main text.

Second, the \(\Pi_v\)-independent part of the \(h^2\)-sector gives
\begin{equation}
\Delta=-\frac12,
\label{eq:AD-Delta-min}
\end{equation}
or, equivalently,
\begin{equation}
\delta_{\mathrm{Tolman}}^{(\mathrm{pressure},\,\mathrm{planar})}
=
-\,\frac{\gamma^\infty}{2}\,\beta_{T,l}.
\label{eq:AD-delta-planar}
\end{equation}

Equation~\eqref{eq:AD-delta-planar} is therefore the planar-limit coefficient obtained within the minimal pressure-side formulation.
It is not a supersaturation-dependent finite-curvature continuation, and it should be distinguished from the retained relations discussed next.

\subsection{First retained pressure-side Tolman-sensitive relation}
\label{app:D:retained}

We next retain the first mixed pressure-side structure produced directly by the symbolic workflow.
At this level, explicit supersaturation dependence is already present, because the liquid-side bivariate expansion generates mixed terms containing both \(\Delta P_v\) and \(\Delta P\).
After substitution of Eq.~\eqref{eq:AD-cap} into Eq.~\eqref{eq:AD-local-balance} and reduction to the dimensionless variables~\eqref{eq:AD-dimless}, one obtains
\begin{equation}
\begin{split}
0
&=
h(-1+\Pi_v)
-
\Pi_v
+
\frac{\Pi_v}{r_\rho}
\\
&\quad
+
h^2
\left[
\frac12+\Delta-\Pi_v+\frac{A_l}{2}\Pi_v-\Delta\Pi_v
\right].
\end{split}
\label{eq:AD-retained-balance}
\end{equation}

Equation~\eqref{eq:AD-retained-balance} is the first retained pressure-side mixed relation.
Its leading retained \(h^0+h^1\)-sector gives the corresponding supersaturation relation
\begin{equation}
\Pi_v
=
\frac{h\,r_\rho}{1+(-1+h)r_\rho},
\label{eq:AD-Piv-retained}
\end{equation}
whereas the retained Tolman-sensitive part yields
\begin{equation}
\Delta
=
\frac{1+(-2+A_l)\Pi_v}{2(-1+\Pi_v)}.
\label{eq:AD-Delta-retained}
\end{equation}

Equation~\eqref{eq:AD-Delta-retained} is not the planar-limit coefficient.
It is a finite-curvature expression with explicit supersaturation dependence.
Its small-\(\Pi_v\) expansion is
\begin{equation}
\begin{split}
\Delta
&=
-\frac12
-
\frac{A_l-1}{2}\Pi_v
+
O(\Pi_v^2)
\\
&=
-\frac12
-
\frac{\beta'_{T,l}}{2\beta_{T,l}^2}\Pi_v
+
O(\Pi_v^2),
\end{split}
\label{eq:AD-Delta-smallPi}
\end{equation}
which in dimensional form becomes
\begin{equation}
\delta_{\mathrm{Tolman}}
=
-\frac{\gamma^\infty}{2}\beta_{T,l}
-
\frac{\gamma^\infty}{2}\beta'_{T,l}\Delta P_v
+
O(\Delta P_v^2).
\label{eq:AD-delta-smallDPv}
\end{equation}

Thus, within the pressure-side expansion, \(\beta'_{T,l}\) first appears once explicit supersaturation dependence is retained.
At the same time, the planar limit of Eq.~\eqref{eq:AD-Delta-retained} is immediate:
\begin{equation}
\Pi_v\to0
\quad\Longrightarrow\quad
\Delta\to-\frac12,
\label{eq:AD-planar-from-retained}
\end{equation}
which reproduces Eq.~\eqref{eq:AD-delta-planar}.

\subsection{Including the next retained supersaturation-dependent terms}
\label{app:D:higher}

If one retains the next \(\Pi_v^2\)-level contributions in the pressure-side algebra, the mixed balance acquires both quadratic vapor-response terms and higher liquid-side mixed terms.
At this level, the Tolman-sensitive relation may still be written in compact form as
\begin{equation}
\Delta
=
\frac{\mathcal{N}_{\mathrm{p}}(\Pi_v)}
     {\mathcal{D}_{\mathrm{p}}(\Pi_v)},
\label{eq:AD-Delta-higher-compact}
\end{equation}
with
\begin{equation}
\begin{split}
\mathcal{N}_{\mathrm{p}}(\Pi_v)
&=
\Bigl[
2+2(-2+A_l)\Pi_v+(4-3A_l)\Pi_v^2
\Bigr]\beta_{T,l}^3
\\
&\quad
+
\Pi_v^2\beta''_{T,l},
\end{split}
\label{eq:AD-Np}
\end{equation}
and
\begin{equation}
\mathcal{D}_{\mathrm{p}}(\Pi_v)
=
2\Bigl[
-2+2\Pi_v+(-2+A_l)\Pi_v^2
\Bigr]\beta_{T,l}^3.
\label{eq:AD-Dp}
\end{equation}

Its small-\(\Pi_v\) expansion is
\begin{equation}
\Delta
=
-\frac12
-
\frac{\beta'_{T,l}}{2\beta_{T,l}^2}\Pi_v
-
\frac{\beta''_{T,l}}{4\beta_{T,l}^3}\Pi_v^2
+
O(\Pi_v^3).
\label{eq:AD-Delta-higher-expand}
\end{equation}
which gives, in dimensional variables,
\begin{equation}
\begin{split}
\delta_{\mathrm{Tolman}}
&=
-\frac{\gamma^\infty}{2}\beta_{T,l}
-
\frac{\gamma^\infty}{2}\beta'_{T,l}\Delta P_v
\\
&\quad
-
\frac{\gamma^\infty}{4}\beta''_{T,l}(\Delta P_v)^2
+
O(\Delta P_v^3).
\end{split}
\label{eq:AD-delta-higher-expand}
\end{equation}

At the same level, the supersaturation branch also acquires explicit vapor-response dependence.
One useful feature of the pressure-side expansion is therefore that it makes transparent the stage at which additional liquid and vapor response derivatives first enter.
At the level of the minimal pressure-side planar-limit coefficient~\eqref{eq:AD-delta-planar}, neither \(\beta'_{T,l}\) nor \(\beta''_{T,l}\) appears.
They arise only in supersaturation-dependent finite-curvature continuations such as Eqs.~\eqref{eq:AD-delta-smallDPv} and~\eqref{eq:AD-delta-higher-expand}.

\subsection{Practical scope of the pressure-side formulation}
\label{app:D:scope}

The pressure-side formulation developed here is one legitimate local formulation of the same capillary--chemical balance used in the main text.
Its main value is structural: it gives a direct pressure-side view of the leading supersaturation relation, the planar-limit coefficient obtained within the minimal pressure-side formulation, and the way in which explicit supersaturation dependence generates finite-curvature continuations involving higher response derivatives.

At the same time, this is not the formulation adopted in the manuscript for the reported planar-limit result relevant to weakly compressible water.
For water near ambient coexistence, the coexistence pressure \(P_0=P_{\mathrm{sat}}(T)\) is typically only of the order of a few kPa, whereas nanometric Laplace overpressures are much larger.
In that regime, an expansion organized directly in excess pressures is usually not the practically controlled one, even though the liquid may still remain weakly compressible in the sense relevant for the adopted density-based formulation.

For that reason, the main text uses the pressure-side formulation only as an alternative local formulation of the same balance, while the reported planar-limit result is taken from the adopted density-based derivation developed in Appendix~\ref{app:E}.
The present Appendix should therefore be read as a pressure-side companion to Sec.~\ref{sec:theory:pressure}, not as the primary formulation used for the manuscript-level planar-limit extraction.
\section{Adopted density-based expansion of the capillary--chemical balance, planar-limit extraction, and comparison variants}
\label{app:E}

This Appendix contains the main density-based derivation used in the manuscript.
It expands the capillary--chemical balance in relative density deviations, specifies the adopted asymmetric retained ordering, derives the corresponding mixed finite-curvature relation, and then extracts the planar-limit coefficient of the Tolman length through the retained Tolman-sensitive sector.
For completeness, the Appendix also records comparison variants, whose role is not to modify the adopted derivation but to clarify how different retained nonlinear contributions lead to different planar coefficients and how one of them recovers the minimal pressure-side planar coefficient obtained in the pressure-side formulation.

\subsection{Density-side setup and adopted retained ordering}
\label{app:E:setup}

Throughout this Appendix, temperature \(T\) is fixed and the reference state is the planar coexistence point \((T,P_0)\), where
\begin{equation}
P_0=P_{\mathrm{sat}}(T),
\qquad
\mu_l(T,P_0)=\mu_v(T,P_0).
\label{eq:AE-coex}
\end{equation}
We use the same excess-pressure variables as in the main text and in Appendix~\ref{app:D}:
\begin{equation}
\Delta P_v\equiv P_v-P_0,
\qquad
\Delta P_l\equiv P_l-P_0,
\qquad
\Delta P\equiv P_l-P_v,
\label{eq:AE-defs}
\end{equation}
so that
\begin{equation}
\Delta P_l=\Delta P_v+\Delta P.
\label{eq:AE-dPl}
\end{equation}
With the sign convention adopted in Sec.~\ref{sec:theory:setup}, this corresponds to a liquid droplet or liquid filament surrounded by vapor, so that the liquid pressure exceeds the vapor pressure by the Laplace amount.

For each phase \(i\in\{l,v\}\), we also introduce the relative density deviation
\begin{equation}
\Delta\rho_i\equiv\frac{\rho_i-\rho_{i0}}{\rho_{i0}},
\qquad
\rho_{i0}\equiv \rho_i(T,P_0),
\label{eq:AE-drho}
\end{equation}
and we write
\begin{equation}
\begin{aligned}
\beta_{T,i} &\equiv \beta_{T,i}(T,P_0),
\\
\beta'_{T,i} &\equiv
\left(\frac{\partial \beta_{T,i}}{\partial P}\right)_T\Bigg|_{P_0}.
\end{aligned}
\label{eq:AE-beta}
\end{equation}

The adopted density-based approximation used in the manuscript combines three specific ingredients:
the leading two-phase density coupling implied by chemical equilibrium,
the one-phase inverse pressure--density expansion retained to second order for the liquid branch,
and only the leading pressure response retained for the vapor branch.
This is the adopted asymmetric formulation used for the manuscript-level derivation.
It is asymmetric not because the liquid and vapor phases obey different equilibrium conditions, but because different local retained orders are used on the two one-phase branches.

\subsection{Leading two-phase density coupling from chemical-potential equality}
\label{app:E:coupling}

The local density-side expansion of the chemical potential obtained in Appendix~\ref{app:B} gives, for each phase \(i\in\{l,v\}\),
\begin{equation}
\mu_i\!\bigl(\rho_{i0}(1+\Delta\rho_i)\bigr)
=
\mu_i(\rho_{i0})
+
\frac{\Delta\rho_i}{\beta_{T,i}\rho_{i0}}
+
O(\Delta\rho_i^2).
\label{eq:AE-mu-linear}
\end{equation}
Imposing chemical equilibrium,
\begin{equation}
\mu_l=\mu_v,
\label{eq:AE-chem}
\end{equation}
and using Eq.~\eqref{eq:AE-coex}, we obtain at leading order
\begin{equation}
\frac{\Delta\rho_l}{\beta_{T,l}\rho_{l0}}
-
\frac{\Delta\rho_v}{\beta_{T,v}\rho_{v0}}
=
0,
\label{eq:AE-leading-bivar}
\end{equation}
or equivalently
\begin{equation}
\Delta\rho_v
=
\Lambda\,\Delta\rho_l
+
O(\Delta\rho_l^2),
\qquad
\Lambda\equiv
\frac{\beta_{T,v}\rho_{v0}}{\beta_{T,l}\rho_{l0}}.
\label{eq:AE-linear-density-coupling}
\end{equation}

This is the leading two-phase density coupling used in the adopted derivation.
At this stage, \(\Delta\rho_l\) and \(\Delta\rho_v\) are still independent one-phase deviations whose coupling is supplied only by chemical equilibrium.
No capillary input and no pressure-side re-expansion has yet been imposed.

\subsection{One-phase liquid and vapor re-expansions used in the adopted derivation}
\label{app:E:reexpansion}

From Appendix~\ref{app:C}, the one-phase inverse excess-pressure / relative-density series gives, for the liquid branch,
\begin{equation}
\Delta\rho_l
=
\beta_{T,l}\Delta P_l
+
\frac12
\left(
1+\frac{\beta'_{T,l}}{\beta_{T,l}^2}
\right)
(\beta_{T,l}\Delta P_l)^2
+
O(\Delta P_l^3).
\label{eq:AE-liquid-inverse}
\end{equation}
For compactness, we define
\begin{equation}
A_l\equiv 1+\frac{\beta'_{T,l}}{\beta_{T,l}^2},
\label{eq:AE-Al}
\end{equation}
so that Eq.~\eqref{eq:AE-liquid-inverse} becomes
\begin{equation}
\Delta\rho_l
=
\beta_{T,l}\Delta P_l
+
\frac12\,A_l(\beta_{T,l}\Delta P_l)^2
+
O(\Delta P_l^3).
\label{eq:AE-liquid-inverse-Al}
\end{equation}

For the vapor branch, the adopted asymmetric derivation retains only the leading pressure response,
\begin{equation}
\Delta\rho_v
=
\beta_{T,v}\Delta P_v
+
O(\Delta P_v^2).
\label{eq:AE-vapor-linear}
\end{equation}

A crucial point is that these are one-phase local re-expansions.
They do not by themselves encode capillary balance or two-phase closure.
Those enter only when Eqs.~\eqref{eq:AE-linear-density-coupling}, \eqref{eq:AE-liquid-inverse-Al}, and \eqref{eq:AE-vapor-linear} are combined.

\subsection{Reduced excess-pressure balance of the adopted density-based formulation}
\label{app:E:balance}

Substituting Eq.~\eqref{eq:AE-liquid-inverse-Al} into the leading density coupling~\eqref{eq:AE-linear-density-coupling} gives
\begin{equation}
\begin{split}
\Delta\rho_v
&=
\frac{\beta_{T,v}\rho_{v0}}{\rho_{l0}}\Delta P_l
+
\frac12\,A_l
\frac{\beta_{T,v}\beta_{T,l}\rho_{v0}}{\rho_{l0}}
\Delta P_l^2
+
O(\Delta P_l^3).
\end{split}
\label{eq:AE-rhov-from-liquid}
\end{equation}
Equating this with the retained vapor-side relation~\eqref{eq:AE-vapor-linear} and dividing by \(\beta_{T,v}\), we obtain
\begin{equation}
\Delta P_v\frac{\rho_{l0}}{\rho_{v0}}
=
\Delta P_l
+
\frac12\,A_l\,\beta_{T,l}\Delta P_l^2
+
O(\Delta P_l^3,\Delta P_v^2).
\label{eq:AE-density-balance}
\end{equation}

Equation~\eqref{eq:AE-density-balance} is the reduced excess-pressure balance specific to the adopted density-based formulation.
It is not a general identity of the capillary--chemical balance.
Rather, it is the particular retained balance obtained when one keeps the leading two-phase density coupling, the quadratic inverse liquid branch, and only the leading vapor pressure response.

\subsection{Mixed finite-curvature balance in \texorpdfstring{$(\Pi_v,h,\Delta)$}{(Pi\_v,h,Delta)}}
\label{app:E:mixed}

We now combine Eq.~\eqref{eq:AE-density-balance} with the Tolman-corrected capillary relation
\begin{equation}
\Delta P_l-\Delta P_v
=
H\gamma^\infty\bigl(1-H\delta_{\mathrm{Tolman}}\bigr).
\label{eq:AE-capillary}
\end{equation}
For convenience, we introduce the same dimensionless variables as in Appendix~\ref{app:D},
\begin{equation}
\begin{aligned}
r_\rho &\equiv \frac{\rho_{v0}}{\rho_{l0}},
&
\Pi_v &\equiv \beta_{T,l}\Delta P_v,
\\
h &\equiv H\gamma^\infty\beta_{T,l},
&
\Delta &\equiv \frac{\delta_{\mathrm{Tolman}}}{\gamma^\infty\beta_{T,l}}.
\end{aligned}
\label{eq:AE-dimless}
\end{equation}
Equation~\eqref{eq:AE-capillary} then implies
\begin{equation}
\beta_{T,l}\Delta P_l
=
\Pi_v+h(1-h\Delta).
\label{eq:AE-liquid-pressure-dimless}
\end{equation}

Substituting Eq.~\eqref{eq:AE-liquid-pressure-dimless} into Eq.~\eqref{eq:AE-density-balance}, multiplying by \(2r_\rho\), and retaining exactly the terms kept by the adopted derivation, one obtains
\begin{equation}
\begin{split}
0
&=
\Pi_v\!\left[-2+2r_\rho+r_\rho A_l\Pi_v\right]
+
2hr_\rho\!\left[1+A_l\Pi_v\right]
\\
&\quad
+
h^2r_\rho\!\left[A_l-2\Delta-2A_l\Pi_v\Delta\right].
\end{split}
\label{eq:AE-implicit-balance}
\end{equation}

At this stage, \(\Pi_v\) is still an independent unknown.
Equation~\eqref{eq:AE-implicit-balance} is therefore a mixed relation in \((\Pi_v,h,\Delta)\), not yet a genuine series in \(h\) alone.
The dependence \(\Pi_v(h)\) is introduced only at the next step as an auxiliary supersaturation-control relation.

This distinction is essential.
The formal status of the retained terms is fixed before any later supersaturation ansatz is imposed.
In particular, the term \(h^2\Pi_v\Delta\) belongs to the explicitly retained \(h^2\)-sector and is not to be reclassified retroactively after a later estimate for \(\Pi_v(h)\) is introduced.

\subsection{Auxiliary supersaturation control relation}
\label{app:E:supersat}

The auxiliary supersaturation-control relation is built only from the explicitly retained \(\Delta\)-free terms of order \(h^0\) and \(h^1\) in Eq.~\eqref{eq:AE-implicit-balance}:
\begin{equation}
\Pi_v\!\left[-2+2r_\rho+r_\rho A_l\Pi_v\right]
+
2hr_\rho\!\left[1+A_l\Pi_v\right]
=
0.
\label{eq:AE-supersat-control}
\end{equation}
This relation is used only as an auxiliary relation controlling the supersaturation.
It is not itself the Tolman-sensitive relation.

Equation~\eqref{eq:AE-supersat-control} is quadratic in \(\Pi_v\):
\begin{equation}
A_l r_\rho \Pi_v^2
+
2\bigl(r_\rho-1+A_lhr_\rho\bigr)\Pi_v
+
2hr_\rho
=
0.
\label{eq:AE-supersat-control-quad}
\end{equation}
The branch satisfying \(\Pi_v\to0\) as \(h\to0\) is
\begin{equation}
\begin{split}
\Pi_v^{\mathrm{phys}}(h)
&=
\frac{
1-r_\rho-A_l h r_\rho
-
\sqrt{
(1-r_\rho)^2
-2A_l h r_\rho
+
A_l^2 h^2 r_\rho^2
}
}{
A_l r_\rho
}.
\end{split}
\label{eq:AE-Pi-branch}
\end{equation}
Its small-\(h\) expansion is
\begin{equation}
\Pi_v
=
\frac{r_\rho}{1-r_\rho}\,h
+
O(h^2),
\label{eq:AE-Pi-leading}
\end{equation}
that is,
\begin{equation}
\Delta P_v
=
H\gamma^\infty\,
\frac{\rho_{v0}}{\rho_{l0}-\rho_{v0}}
+
O(H^2).
\label{eq:AE-supersat-dim}
\end{equation}
which reproduces the leading Kelvin-type scaling quoted in the main text.

\subsection{Tolman-sensitive relation and planar-limit extraction}
\label{app:E:strict}

After the auxiliary supersaturation step has fixed the leading scaling of \(\Pi_v\), the Tolman contribution is extracted only from the explicitly retained \(h^2\)-sector of Eq.~\eqref{eq:AE-implicit-balance}.
That sector is
\begin{equation}
A_l-2\Delta-2A_l\Pi_v\Delta=0,
\label{eq:AE-h2-sector}
\end{equation}
which yields the Tolman-sensitive relation
\begin{equation}
\Delta(\Pi_v)
=
\frac{A_l}{2\left(1+A_l\Pi_v\right)}.
\label{eq:AE-Delta-Piv}
\end{equation}

The planar-limit coefficient is obtained only afterward, by taking the planar limit of this retained Tolman-sensitive relation:
\begin{equation}
\Delta_0
=
\lim_{\Pi_v\to0}\Delta(\Pi_v)
=
\frac{A_l}{2}
=
\frac12\left(1+\frac{\beta'_{T,l}}{\beta_{T,l}^2}\right).
\label{eq:AE-Delta0}
\end{equation}
Returning to dimensional variables gives
\begin{equation}
\delta_{\mathrm{Tolman}}^{\mathrm{planar}}
=
\frac{\gamma^\infty}{2}
\left(
\beta_{T,l}
+
\frac{\beta'_{T,l}}{\beta_{T,l}}
\right).
\label{eq:AE-delta-planar}
\end{equation}

Here \(\delta_{\mathrm{Tolman}}^{\mathrm{planar}}\) is the planar-limit value of the same Tolman length \(\delta_{\mathrm{Tolman}}\), not a separate parameter introduced for a strictly planar interface.
The planar-limit consequence of the adopted density-based derivation is therefore the coefficient extracted in Eq.~\eqref{eq:AE-delta-planar}, whereas the finite-\(\Pi_v\) relation~\eqref{eq:AE-Delta-Piv} remains a retained continuation.

\subsection{Finite-curvature continuation}
\label{app:E:continuation}

If one keeps Eq.~\eqref{eq:AE-Delta-Piv} at finite \(\Pi_v\), rather than taking the strict planar limit immediately, one obtains the non-strict finite-curvature continuation
\begin{equation}
\Delta_{\mathrm{cont}}(\Pi_v)
=
\frac{A_l}{2\left(1+A_l\Pi_v\right)}.
\label{eq:AE-Delta-cont}
\end{equation}
In dimensional form,
\begin{equation}
\delta_{\mathrm{Tolman}}^{\mathrm{cont}}
=
\frac{
\delta_{\mathrm{Tolman}}^{\mathrm{planar}}
}{
1+
\left(
\beta_{T,l}+\frac{\beta'_{T,l}}{\beta_{T,l}}
\right)\Delta P_v
}.
\label{eq:AE-delta-cont}
\end{equation}

This expression reduces to the planar-limit result as \(\Delta P_v\to0\), but it is not itself the planar-limit coefficient of the retained curvature expansion.
Its status is therefore that of a retained finite-curvature continuation.

\subsection{Decomposition of the quadratic liquid contribution under composition of local series}
\label{app:E:decomposition}

To understand why comparison variants can yield different planar coefficients, it is useful to isolate the two distinct quadratic liquid contributions before they are composed.

From Appendix~\ref{app:B}, the local liquid chemical-potential expansion in the relative density deviation may be written as
\begin{equation}
\Delta\mu_l
=
a_1\Delta\rho_l+a_2\Delta\rho_l^2+O(\Delta\rho_l^3),
\label{eq:AE-mu-density-local}
\end{equation}
with
\begin{equation}
a_1=\frac{1}{\beta_{T,l}\rho_{l0}},
\qquad
a_2=-\frac{(1+A_l)}{2\beta_{T,l}\rho_{l0}}.
\label{eq:AE-a1a2}
\end{equation}
From Appendix~\ref{app:C}, the inverse one-phase liquid series is
\begin{equation}
\Delta\rho_l
=
b_1\Delta P_l+b_2\Delta P_l^2+O(\Delta P_l^3),
\label{eq:AE-rho-pressure-local}
\end{equation}
with
\begin{equation}
b_1=\beta_{T,l},
\qquad
b_2=\frac12\,A_l\beta_{T,l}^2.
\label{eq:AE-b1b2}
\end{equation}

Composing Eqs.~\eqref{eq:AE-mu-density-local} and~\eqref{eq:AE-rho-pressure-local} gives
\begin{equation}
\Delta\mu_l
=
a_1b_1\Delta P_l
+
\left(a_1b_2+a_2b_1^2\right)\Delta P_l^2
+
O(\Delta P_l^3).
\label{eq:AE-composed-general}
\end{equation}
The linear coefficient is
\begin{equation}
a_1b_1=\frac{1}{\rho_{l0}},
\label{eq:AE-composed-linear}
\end{equation}
whereas the quadratic coefficient splits into two distinct pieces,
\begin{equation}
a_1b_2
=
\frac{A_l\beta_{T,l}}{2\rho_{l0}},
\qquad
a_2b_1^2
=
-\frac{(1+A_l)\beta_{T,l}}{2\rho_{l0}}.
\label{eq:AE-composed-split}
\end{equation}
Their sum is
\begin{equation}
a_1b_2+a_2b_1^2
=
-\frac{\beta_{T,l}}{2\rho_{l0}}.
\label{eq:AE-composed-sum}
\end{equation}

Thus, the quadratic liquid contribution in the pressure representation is not primitive.
It emerges only after composition of two local series.
For this reason, different retained constructions may keep only one of the two quadratic channels, or both together, and this is the algebraic source of the comparison coefficients recorded below.

\subsection{Comparison variants and coefficient migration}
\label{app:E:comparison}

We now record three comparison variants, distinguished by which quadratic liquid contribution survives in the retained \(h^2\)-sector after composition.

\paragraph*{(i) Chemical-potential quadratic contribution only.}
If one retains only the quadratic contribution originating from the local chemical-potential expansion, the representative reduced balance may be written as
\begin{equation}
\frac{\Delta P_l}{\rho_{l0}}
-
\frac{\Delta P_v}{\rho_{v0}}
-
\frac{(1+A_l)\beta_{T,l}}{2\rho_{l0}}\Delta P_l^2
=
0.
\label{eq:AE-comp-mu-balance}
\end{equation}
The corresponding Tolman-sensitive relation is
\begin{equation}
\Delta(\Pi_v)
=
\frac{1+A_l}{2\left(-1+\Pi_v+A_l\Pi_v\right)},
\label{eq:AE-comp-mu-DeltaPiv}
\end{equation}
and therefore
\begin{equation}
\Delta_0^{(\mu)}
=
\lim_{\Pi_v\to0}\Delta(\Pi_v)
=
-\frac{1+A_l}{2}.
\label{eq:AE-comp-mu-planar}
\end{equation}

\paragraph*{(ii) Inverse pressure--density quadratic contribution only.}
If one retains only the quadratic contribution coming from the inverse one-phase liquid series, the representative reduced balance becomes
\begin{equation}
\frac{\Delta P_l}{\rho_{l0}}
-
\frac{\Delta P_v}{\rho_{v0}}
+
\frac{A_l\beta_{T,l}}{2\rho_{l0}}\Delta P_l^2
=
0.
\label{eq:AE-comp-inv-balance}
\end{equation}
Then
\begin{equation}
\Delta(\Pi_v)
=
\frac{A_l}{2\left(1+A_l\Pi_v\right)},
\label{eq:AE-comp-inv-DeltaPiv}
\end{equation}
so that
\begin{equation}
\Delta_0^{(\mathrm{inv})}
=
\lim_{\Pi_v\to0}\Delta(\Pi_v)
=
\frac{A_l}{2}.
\label{eq:AE-comp-inv-planar}
\end{equation}
This is precisely the planar coefficient produced by the adopted density-based derivation used in the manuscript.

\paragraph*{(iii) Both quadratic liquid contributions retained together.}
If both quadratic liquid contributions are retained in the same comparison construction, the resulting Tolman-sensitive relation takes the form
\begin{equation}
\Delta(\Pi_v)
=
\frac{\mathcal{N}_{\mu+\mathrm{inv}}(\Pi_v)}
     {\mathcal{D}_{\mu+\mathrm{inv}}(\Pi_v)},
\label{eq:AE-comp-both-DeltaPiv}
\end{equation}
where
\[
\mathcal{N}_{\mu+\mathrm{inv}}(\Pi_v)
=
2
+
6A_l\Pi_v
+
3A_l^3\Pi_v^2
+
3A_l^2\Pi_v(2+\Pi_v),
\]
and
\[
\mathcal{D}_{\mu+\mathrm{inv}}(\Pi_v)
=
-4
+
4\Pi_v
+
6A_l(1+A_l)\Pi_v^2.
\]
Hence,
\begin{equation}
\Delta_0^{(\mu+\mathrm{inv})}
=
\lim_{\Pi_v\to0}\Delta(\Pi_v)
=
-\frac12.
\label{eq:AE-comp-both-planar}
\end{equation}

Returning to dimensional variables, the three comparison coefficients are
\begin{align}
\delta_{\mathrm{Tolman}}^{(\mu,\mathrm{planar})}
=
-\frac{\gamma^\infty\beta_{T,l}}{2}(1+A_l)&=
-\gamma^\infty\beta_{T,l}
-
\frac{\gamma^\infty}{2}\,
\frac{\beta'_{T,l}}{\beta_{T,l}},
\label{eq:AE-comp-mu-dim}
\\
\delta_{\mathrm{Tolman}}^{(\mathrm{inv},\mathrm{planar})}
=
\frac{\gamma^\infty\beta_{T,l}}{2}A_l
&=
\frac{\gamma^\infty}{2}
\left(
\beta_{T,l}+\frac{\beta'_{T,l}}{\beta_{T,l}}
\right),
\label{eq:AE-comp-inv-dim}
\\
\delta_{\mathrm{Tolman}}^{(\mu+\mathrm{inv},\mathrm{planar})}
&=
-\frac{\gamma^\infty\beta_{T,l}}{2}.
\label{eq:AE-comp-both-dim}
\end{align}

The coefficient migration across these comparison constructions is therefore controlled by which quadratic liquid contribution survives in the retained \(h^2\)-sector.
The comparison variants do not alter the adopted derivation; they only classify the alternative coefficients generated by different retained nonlinear patterns.

\subsection{Recovery of the pressure-side planar coefficient}
\label{app:E:coincidence}

A noteworthy feature of the comparison variants is that the coefficient obtained when both quadratic liquid contributions are retained together,
\begin{equation}
\Delta_0^{(\mu+\mathrm{inv})}=-\frac12,
\label{eq:AE-pressure-coincidence}
\end{equation}
recovers the minimal pressure-side planar coefficient derived in the pressure-side formulation of Appendix~\ref{app:D}.
In dimensional form,
\begin{equation}
\delta_{\mathrm{Tolman}}^{(\mu+\mathrm{inv},\mathrm{planar})}
=
-\frac{\gamma^\infty\beta_{T,l}}{2}
=
\delta_{\mathrm{Tolman}}^{(\mathrm{pressure},\,\mathrm{planar})}.
\label{eq:AE-pressure-coincidence-dim}
\end{equation}

This recovery is an expected algebraic consequence of composing the local density expansion with the inverse pressure--density series to the same retained order.
In that case, the relative density deviation acts only as an intermediate parametrization and the corresponding pressure-side coefficient is recovered.
It does not imply that the adopted density-based derivation and the pressure-side derivation are equivalent.
The two formulations remain different local organizations of the same capillary--chemical balance, with different retained variables and different control conditions.

\subsection{Status of the comparison variants}
\label{app:E:status}

The adopted manuscript-level derivation is the one given in Secs.~\ref{app:E:setup}--\ref{app:E:continuation}.
Within that derivation, the reported planar-limit result is Eq.~\eqref{eq:AE-delta-planar}.
The comparison variants introduced afterward do not modify that result and are not alternative adopted formulations.

Their function is narrower.
They document how different retained nonlinear contributions can migrate into the retained \(h^2\)-sector and thereby generate different planar coefficients.
They also clarify why retaining both quadratic liquid channels recovers the minimal pressure-side planar coefficient and why the adopted density-based coefficient corresponds to the asymmetric retained ordering used for the weakly compressible water regime emphasized in the manuscript.

Accordingly, the comparison variants should be read as algebraic classification and interpretation tools, not as competing manuscript-level derivations.
\section{Water-specific hierarchy support for the adopted density-based mixed balance}
\label{app:F}

This Appendix provides water-specific support for the adopted density-based mixed-balance formulation using the IAPWS-IF97 industrial formulation.
Its purpose is not to derive a new formulation, but to check \textit{a posteriori} that, for water in the regime emphasized in the manuscript, the explicit terms retained at order \(h^2\) remain sufficiently small relative to the explicit \(\Delta\)-free terms of orders \(h^0\) and \(h^1\).
In this way, the Appendix supports the practical relevance of the adopted density-based formulation for water while remaining logically separate from the derivation itself.

\subsection{Purpose and scope}
\label{app:F:purpose}

The adopted density-based derivation has already been completed in Appendix~\ref{app:E}.
In particular, the auxiliary supersaturation-control relation is constructed there only from the explicitly retained \(\Delta\)-free terms of orders \(h^0\) and \(h^1\), whereas the Tolman-sensitive relation is extracted only from the explicitly retained sector of order \(h^2\).
The present Appendix does not alter that logic, does not replace it by a different asymptotic formulation, and does not introduce a new approximation pattern.

Its role is narrower.
For water, and specifically for the weakly compressible near-coexistence regime emphasized in the manuscript, we ask whether the explicit sector retained at order \(h^2\) remains sufficiently small relative to the explicit \(\Delta\)-free sector of orders \(h^0\) and \(h^1\) over the practically relevant nanometric range.
If this hierarchy is supported by an independent EOS-based evaluation, then the adopted formulation of Appendix~\ref{app:E} gains additional water-specific practical support.

Accordingly, this Appendix should be read as a water-specific diagnostic appendix.
It supports the practical regime relevance of the adopted density-based mixed balance, but it is not part of the logical derivation of the planar-limit coefficient itself.

\subsection{Diagnostic quantities inherited from Appendix~\ref{app:E}}
\label{app:F:diagnostics}

We begin from the retained mixed balance derived in Appendix~\ref{app:E},
\begin{equation}
\begin{split}
0
&=
\Pi_v\!\left[-2+2r_\rho+r_\rho A_l\Pi_v\right]
+
2hr_\rho\!\left[1+A_l\Pi_v\right]
\\
&\quad
+
h^2r_\rho\!\left[A_l-2\Delta-2A_l\Pi_v\Delta\right].
\end{split}
\label{eq:AF-implicit-balance}
\end{equation}
Here
\begin{equation}
\begin{aligned}
r_\rho &\equiv \frac{\rho_{v0}}{\rho_{l0}},
&
\Pi_v &\equiv \beta_{T,l}\Delta P_v,
\\
h &\equiv H\gamma^\infty\beta_{T,l},
&
\Delta &\equiv \frac{\delta_{\mathrm{Tolman}}}{\gamma^\infty\beta_{T,l}},
\end{aligned}
\label{eq:AF-dimless}
\end{equation}
and
\begin{equation}
A_l\equiv 1+\frac{\beta'_{T,l}}{\beta_{T,l}^2}.
\label{eq:AF-Al}
\end{equation}

For the present diagnostic purpose, it is convenient to separate Eq.~\eqref{eq:AF-implicit-balance} into two parts:
\begin{equation}
\mathcal{S}_{1}(\Pi_v,h)+\mathcal{S}_2(\Pi_v,h,\Delta)=0,
\label{eq:AF-S01S2}
\end{equation}
where
\begin{equation}
\begin{split}
\mathcal{S}_{1}(\Pi_v,h)
&\equiv
\Pi_v\!\left[-2+2r_\rho+r_\rho A_l\Pi_v\right]
\\
&\quad
+
2h r_\rho\!\left[1+A_l\Pi_v\right].
\end{split}
\label{eq:AF-S01}
\end{equation}
contains exactly the explicitly retained \(\Delta\)-free terms of orders \(h^0\) and \(h^1\), whereas
\begin{equation}
\mathcal{S}_2(\Pi_v,h,\Delta)
=
h^2r_\rho\!\left[A_l-2\Delta-2A_l\Pi_v\Delta\right]
\label{eq:AF-S2}
\end{equation}
contains the explicitly retained sector of order \(h^2\).

This decomposition is used here only for hierarchy diagnostics.
It must not be interpreted as a retroactive reorganization of the adopted mixed balance.
In particular, the term \(h^2\Pi_v\Delta\) remains part of the explicitly retained \(h^2\)-sector and is not reclassified afterward on the basis of any later estimate for \(\Pi_v(h)\).

As a dimensionless diagnostic ratio characterizing the relative importance of the retained \(h^2\)-sector, we define
\begin{equation}
\mathcal{R}(T,R,\Delta P_v,\delta_{\mathrm{Tolman}})
\equiv
\frac{\mathcal{S}_2}{\mathcal{S}_{1}}.
\label{eq:AF-R}
\end{equation}
This quantity is used below only as a diagnostic ratio.
For hierarchy assessment, the relevant condition is \(|\mathcal{R}|\ll1\).
Small values of \(|\mathcal{R}|\) indicate that, for the water state under consideration, the explicitly retained \(h^2\)-sector remains subdominant relative to the explicitly retained \(\Delta\)-free sector of orders \(h^0\) and \(h^1\).

\subsection{EOS-based evaluation procedure for water}
\label{app:F:eos}

To evaluate the hierarchy ratio~\eqref{eq:AF-R} for water, we use thermodynamic properties evaluated from the IAPWS-IF97 industrial formulation as empirical input for the required liquid and vapor quantities near coexistence.
Throughout, the coexistence reference state is \((T,P_0)\), with \(P_0=P_{\mathrm{sat}}(T)\), and all coefficients entering Eqs.~\eqref{eq:AF-dimless}--\eqref{eq:AF-S2} are evaluated from the corresponding coexistence state at the chosen temperature.

For a chosen temperature \(T\), curvature radius \(R\), and prescribed Tolman length \(\delta_{\mathrm{Tolman}}\), we first compute the coexistence quantities \(P_0\), \(\rho_{l0}\), \(\rho_{v0}\), \(\beta_{T,l}\), \(\beta'_{T,l}\), and \(\gamma^\infty\). These quantities determine \(r_\rho\), \(A_l\), and the curvature scale \(h\).

A crucial point is that, in the present Appendix, \(\Pi_v\) is not generated from the auxiliary supersaturation-control relation of Appendix~\ref{app:E}.
Instead, the vapor excess pressure \(\Delta P_v\) is obtained independently from the corresponding water-specific capillary-equilibrium calculation and only afterward converted to
\begin{equation}
\Pi_v=\beta_{T,l}\Delta P_v.
\label{eq:AF-Piv}
\end{equation}
Thus, the hierarchy test is not circular: it does not reuse the same auxiliary relation whose practical relevance it is intended to support.

For the geometries considered in the manuscript, the dimensionless curvature parameter is
\begin{equation}
h=\frac{(n-1)\gamma^\infty\beta_{T,l}}{R},
\label{eq:AF-h}
\end{equation}
with \(n=3\) for spheres and \(n=2\) for cylinders.
Once \(h\), \(\Pi_v\), and \(\Delta\) are specified, the quantities \(\mathcal{S}_{1}\), \(\mathcal{S}_2\), and \(\mathcal{R}\) follow directly from Eqs.~\eqref{eq:AF-S01}, \eqref{eq:AF-S2}, and~\eqref{eq:AF-R}.

In this way, the present Appendix uses the adopted mixed balance only as a diagnostic object.
The EOS-based evaluation is external to the derivation itself and serves only to test, for water, whether the retained hierarchy underlying the adopted formulation remains practically plausible over the parameter range of interest.

\subsection{Water-specific hierarchy behavior}
\label{app:F:behavior}

For water away from criticality, and for the nanometric regime emphasized in the manuscript, the EOS-based estimates show that the magnitude of the diagnostic ratio~\eqref{eq:AF-R} remains typically below unity and, over a broad practically relevant range, substantially below unity.
In representative near-ambient conditions, one finds
\begin{equation}
|\mathcal{R}|\sim 10^{-2}\text{--}10^{-1}
\label{eq:AF-R-scale}
\end{equation}
for a wide range of nanometric radii and moderate choices of \(\delta_{\mathrm{Tolman}}\).

Thus, for water in the regime emphasized here, the explicitly retained sector of order \(h^2\) remains typically smaller than the explicitly retained \(\Delta\)-free sector of orders \(h^0\) and \(h^1\) by roughly one to two orders of magnitude.
This is precisely the hierarchy required for the practical usefulness of the adopted mixed-balance formulation.
It does not make the adopted formulation exact, but it does support the claim that the retained \(h^2\)-sector remains practically subdominant over the regime of interest.

\begin{figure*}[t]
\centering
\includegraphicsmaybe[width=\textwidth]{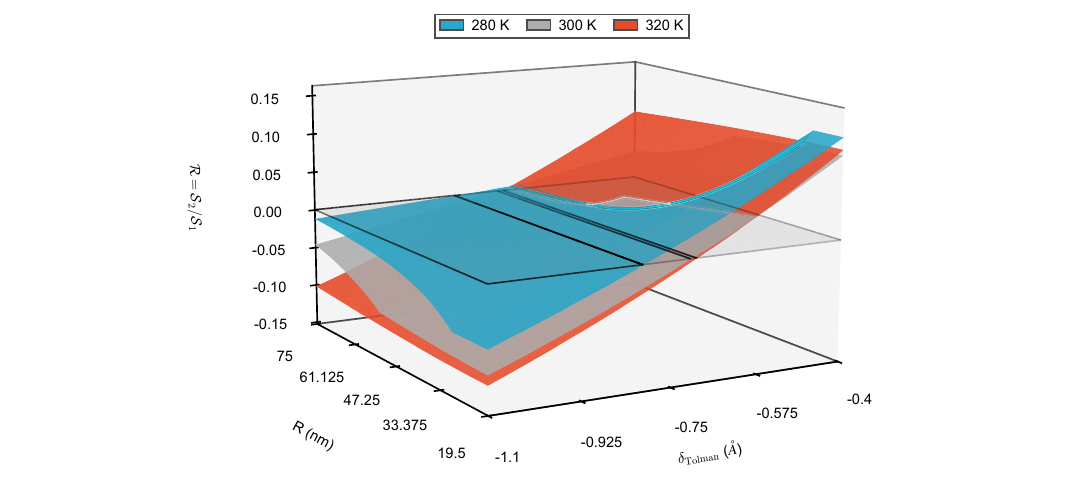}
\caption{\label{fig:R-ratio-water}
Water-specific hierarchy support for the adopted density-based mixed balance.
Shown is the signed ratio \(\mathcal{R}=\mathcal{S}_2/\mathcal{S}_{1}\), where \(\mathcal{S}_{1}\) denotes the explicitly retained \(\Delta\)-free sector of orders \(h^0\) and \(h^1\), and \(\mathcal{S}_2\) denotes the explicitly retained sector of order \(h^2\) in the mixed balance inherited from Appendix~\ref{app:E}.
The thermodynamic input is evaluated for water using the IAPWS-IF97 formulation.
Over the nanometric regime emphasized in the manuscript and for moderate choices of \(\delta_{\mathrm{Tolman}}\), the magnitude \(|\mathcal{R}|\) remains typically at the level \(10^{-2}\text{--}10^{-1}\), supporting the practical subdominance of the retained \(h^2\)-sector relative to the retained \(\Delta\)-free sector.}
\end{figure*}

Figure~\ref{fig:R-ratio-water} illustrates this behavior.
The plotted quantity is the signed ratio \(\mathcal{R}=\mathcal{S}_2/\mathcal{S}_{1}\), evaluated for water from the IAPWS-IF97-based thermodynamic input over the parameter window emphasized in the manuscript.
Its magnitude \(|\mathcal{R}|\) is the quantity relevant for the hierarchy assessment.

The role of this figure is therefore not merely illustrative.
It supports, on the water example, that the retained truncation levels used in the adopted density-based formulation remain practically adequate over the parameter range of interest.
This is why the plot is retained in the present appendix.

\subsection{Practical implication for the adopted density-based formulation}
\label{app:F:implication}

For water under the conditions considered in the manuscript, the hierarchy condition \(|\mathcal{R}|\ll1\) supports the practical relevance of the adopted density-based formulation developed in Appendix~\ref{app:E}.
More specifically, it supports the strategy in which the auxiliary supersaturation-control relation is built from the explicitly retained \(\Delta\)-free sector of orders \(h^0\) and \(h^1\), while the Tolman-sensitive relation is extracted from the explicitly retained sector of order \(h^2\), and only afterward is the planar limit taken.

A convenient water-specific practical criterion is therefore the combined requirement
\begin{equation}
|\mathcal{R}|\ll1,
\qquad
|\beta_{T,l}\Delta P_l|\ll1.
\label{eq:AF-criterion}
\end{equation}
The first condition checks the retained hierarchy inside the adopted mixed balance.
The second checks that the liquid branch remains within the weak-compressibility regime underlying the density-based local re-expansion itself.

It is important to emphasize what this support means and what it does not mean.
It supports the practical regime relevance of the adopted density-based formulation for water.
It does not elevate the auxiliary supersaturation-control relation to the status of an independently proven universal asymptotic law, and it does not modify the logical status of the planar-limit extraction already completed in Appendix~\ref{app:E}.

\subsection{Limitations}
\label{app:F:limits}

The present hierarchy support is intentionally water-specific and regime-specific.
It is aimed at the weakly compressible near-coexistence regime emphasized in the manuscript, not at critical or near-spinodal conditions.
Near the critical region, the coexistence density contrast decreases, the compressibilities grow strongly, and both the local density-based expansion and the hierarchy expressed by \(\mathcal{R}\) can lose practical control.

This Appendix also does not claim that the pressure-side formulation becomes invalid in principle.
Rather, its point is narrower: for water near ambient coexistence, the density-based formulation is the practically more useful one over the nanometric regime emphasized here, and the EOS-based hierarchy estimates support that choice.
In another substance or in another thermodynamic regime, the practically relevant control condition may be different.

Most importantly, the present Appendix must not be read as a second derivation of the manuscript result.
It is only an \textit{a posteriori} diagnostic appendix.
Its sole function is to support, for water, the practical relevance of the adopted separation between the explicitly retained \(\Delta\)-free sector of orders \(h^0\) and \(h^1\) and the explicitly retained Tolman-sensitive sector of order \(h^2\).
\bibliography{aipsamp}

@article{MalijevskyJackson2012,
  author    = {Malijevsk{\'y}, Alexandr and Jackson, George},
  title     = {A perspective on the interfacial properties of nanoscopic liquid drops},
  journal   = {Journal of Physics: Condensed Matter},
  year      = {2012},
  month     = {oct},
  volume    = {24},
  number    = {46},
  pages     = {464121},
  publisher = {IOP Publishing},
  doi       = {10.1088/0953-8984/24/46/464121},
  url       = {https://doi.org/10.1088/0953-8984/24/46/464121}
}

@article{Sampayo2010,
  author  = {Sampayo, Jos{\'e} G. and Malijevsk{\'y}, Alexandr and M{\"u}ller, Erich A. and de Miguel, Enrique and Jackson, George},
  title   = {Communications: Evidence for the role of fluctuations in the thermodynamics of nanoscale drops and the implications in computations of the surface tension},
  journal = {The Journal of Chemical Physics},
  volume  = {132},
  number  = {14},
  pages   = {141101},
  year    = {2010},
  month   = {04},
  issn    = {0021-9606},
  doi     = {10.1063/1.3376612},
  url     = {https://doi.org/10.1063/1.3376612},
  eprint  = {https://pubs.aip.org/aip/jcp/article-pdf/doi/10.1063/1.3376612/16122339/141101\_1\_online.pdf}
}

@article{doi:10.1021/jacs.5c14775,
  author  = {Chen, Bin and Nguyen, Ngoc My Nhi and Johnson, Kirsten K.},
  title   = {Curvature-Corrected Surface Tension Governs Nucleation in Molecular Systems},
  journal = {Journal of the American Chemical Society},
  volume  = {147},
  number  = {47},
  pages   = {43816--43821},
  year    = {2025},
  doi     = {10.1021/jacs.5c14775},
  note    = {PMID: 41247795},
  url     = {https://doi.org/10.1021/jacs.5c14775}
}

@book{Gibbs1878OnTE,
  author    = {Gibbs, J. W.},
  title     = {Scientific Papers of J. Willard Gibbs, in Two Volumes},
  publisher = {Longmans, Green},
  address   = {New York},
  year      = {1906},
  url       = {https://books.google.com.ua/books?id=-neYVEbAm4oC},
  lccn      = {agr07001540}
}

@article{LAMMPS,
  author  = {Thompson, A. P. and Aktulga, H. M. and Berger, R. and Bolintineanu, D. S. and Brown, W. M. and Crozier, P. S. and in 't Veld, P. J. and Kohlmeyer, A. and Moore, S. G. and Nguyen, T. D. and Shan, R. and Stevens, M. J. and Tranchida, J. and Trott, C. and Plimpton, S. J.},
  title   = {{LAMMPS}---a flexible simulation tool for particle-based materials modeling at the atomic, meso, and continuum scales},
  journal = {Comp. Phys. Comm.},
  volume  = {271},
  pages   = {108171},
  year    = {2022},
  doi     = {10.1016/j.cpc.2021.108171},
  url     = {https://doi.org/10.1016/j.cpc.2021.108171}
}

@article{10.1063/1.1747247,
  author  = {Tolman, Richard C.},
  title   = {The Effect of Droplet Size on Surface Tension},
  journal = {The Journal of Chemical Physics},
  volume  = {17},
  number  = {3},
  pages   = {333--337},
  year    = {1949},
  doi     = {10.1063/1.1747247},
  url     = {https://doi.org/10.1063/1.1747247}
}

@article{10.1063/1.1747248,
  author  = {Kirkwood, John G. and Buff, Frank P.},
  title   = {The Statistical Mechanical Theory of Surface Tension},
  journal = {The Journal of Chemical Physics},
  volume  = {17},
  number  = {3},
  pages   = {338--343},
  year    = {1949},
  doi     = {10.1063/1.1747248},
  url     = {https://doi.org/10.1063/1.1747248}
}

@article{e21070670,
  author  = {Schmelzer, J{\"u}rgen W. P. and Abyzov, Alexander S. and Baidakov, Vladimir G.},
  title   = {Entropy and the Tolman Parameter in Nucleation Theory},
  journal = {Entropy},
  volume  = {21},
  number  = {7},
  pages   = {670},
  year    = {2019},
  doi     = {10.3390/e21070670},
  url     = {https://doi.org/10.3390/e21070670}
}

@article{10.1063/1.2167642,
  author  = {Blokhuis, Edgar M. and Kuipers, Joris},
  title   = {Thermodynamic expressions for the Tolman length},
  journal = {The Journal of Chemical Physics},
  volume  = {124},
  number  = {7},
  pages   = {074701},
  year    = {2006},
  doi     = {10.1063/1.2167642},
  url     = {https://doi.org/10.1063/1.2167642}
}

@article{doi:10.1021/jp0108723,
  author  = {Bykov, T. V. and Zeng, X. C.},
  title   = {Statistical Mechanics of Surface Tension and Tolman Length of Dipolar Fluids},
  journal = {The Journal of Physical Chemistry B},
  volume  = {105},
  number  = {47},
  pages   = {11586--11594},
  year    = {2001},
  doi     = {10.1021/jp0108723},
  url     = {https://doi.org/10.1021/jp0108723}
}

@article{10.1063/1.479650,
  author  = {Bykov, T. V. and Zeng, X. C.},
  title   = {A patching model for surface tension and the Tolman length},
  journal = {The Journal of Chemical Physics},
  volume  = {111},
  number  = {8},
  pages   = {3705--3713},
  year    = {1999},
  doi     = {10.1063/1.479650},
  url     = {https://doi.org/10.1063/1.479650}
}

@article{10.1063/1.480434,
  author  = {Bykov, T. V. and Zeng, X. C.},
  title   = {A patching model for surface tension of spherical droplet and Tolman length. II},
  journal = {The Journal of Chemical Physics},
  volume  = {111},
  number  = {23},
  pages   = {10602--10610},
  year    = {1999},
  doi     = {10.1063/1.480434},
  url     = {https://doi.org/10.1063/1.480434}
}

@article{10.1063/1.1354165,
  author  = {Napari, Ismo and Laaksonen, Ari},
  title   = {The effect of potential truncation on the gas--liquid surface tension of planar interfaces and droplets},
  journal = {The Journal of Chemical Physics},
  volume  = {114},
  number  = {13},
  pages   = {5796--5801},
  year    = {2001},
  doi     = {10.1063/1.1354165},
  url     = {https://doi.org/10.1063/1.1354165}
}

@article{doi:10.1021/jp011028f,
  author  = {Bartell, Lawrence S.},
  title   = {Tolman's {$\delta$}, Surface Curvature, Compressibility Effects, and the Free Energy of Drops},
  journal = {The Journal of Physical Chemistry B},
  volume  = {105},
  number  = {47},
  pages   = {11615--11618},
  year    = {2001},
  doi     = {10.1021/jp011028f},
  url     = {https://doi.org/10.1021/jp011028f}
}

@Article{Wang_2013,
  author  = {Wang, Xiao-Song and Zhu, Ru-Zeng},
  title   = {Relation between Tolman length and isothermal compressibility for simple liquids},
  journal = {Chinese Physics B},
  volume  = {22},
  number  = {3},
  pages   = {036801},
  year    = {2013},
  doi     = {10.1088/1674-1056/22/3/036801},
  url     = {https://dx.doi.org/10.1088/1674-1056/22/3/036801}
}

@Article{A.Laaksonen_1996,
  author  = {Laaksonen, A. and McGraw, R.},
  title   = {Thermodynamics, gas-liquid nucleation, and size-dependent surface tension},
  journal = {Europhysics Letters},
  volume  = {35},
  number  = {5},
  pages   = {367},
  year    = {1996},
  doi     = {10.1209/epl/i1996-00121-4},
  url     = {https://dx.doi.org/10.1209/epl/i1996-00121-4}
}

@Article{PhysRevE.95.062801,
  author  = {Burian, Sergii and Isaiev, Mykola and Termentzidis, Konstantinos and Sysoev, Vladimir and Bulavin, Leonid},
  title   = {Size dependence of the surface tension of a free surface of an isotropic fluid},
  journal = {Phys. Rev. E},
  volume  = {95},
  number  = {6},
  pages   = {062801},
  year    = {2017},
  doi     = {10.1103/PhysRevE.95.062801},
  url     = {https://link.aps.org/doi/10.1103/PhysRevE.95.062801}
}

@Article{C1CP22168J,
  author  = {Vega, Carlos and Abascal, Jose L. F.},
  title   = {Simulating water with rigid non-polarizable models: a general perspective},
  journal = {Phys. Chem. Chem. Phys.},
  volume  = {13},
  number  = {44},
  pages   = {19663--19688},
  year    = {2011},
  doi     = {10.1039/C1CP22168J},
  url     = {http://dx.doi.org/10.1039/C1CP22168J}
}

@Article{10.1063/1.2121687,
  author  = {Abascal, J. L. F. and Vega, C.},
  title   = {A general purpose model for the condensed phases of water: TIP4P/2005},
  journal = {The Journal of Chemical Physics},
  volume  = {123},
  number  = {23},
  pages   = {234505},
  year    = {2005},
  doi     = {10.1063/1.2121687},
  url     = {https://doi.org/10.1063/1.2121687}
}

@Article{VegaDeMiguel2007,
  author  = {Vega, C. and de Miguel, E.},
  title   = {Surface tension of the most popular models of water by using the test-area simulation method},
  journal = {The Journal of Chemical Physics},
  volume  = {126},
  number  = {15},
  pages   = {154707},
  year    = {2007},
  doi     = {10.1063/1.2715577},
  url     = {https://doi.org/10.1063/1.2715577}
}

@Inbook{Wagner2008,
  title     = {IAPWS Industrial Formulation 1997 for the Thermodynamic Properties of Water and Steam},
  bookTitle = {International Steam Tables: Properties of Water and Steam Based on the Industrial Formulation IAPWS-IF97},
  year      = {2008},
  publisher = {Springer Berlin Heidelberg},
  address   = {Berlin, Heidelberg},
  pages     = {7--150},
  isbn      = {978-3-540-74234-0},
  doi       = {10.1007/978-3-540-74234-0_3},
  url       = {https://doi.org/10.1007/978-3-540-74234-0_3}
}

@InBook{Berendsen1981,
  author    = {Berendsen, H. J. C. and Postma, J. P. M. and van Gunsteren, W. F. and Hermans, J.},
  title     = {Interaction Models for Water in Relation to Protein Hydration},
  booktitle = {Intermolecular Forces: Proceedings of the Fourteenth Jerusalem Symposium on Quantum Chemistry and Biochemistry Held in Jerusalem, Israel, April 13--16, 1981},
  pages     = {331--342},
  publisher = {Springer Netherlands},
  address   = {Dordrecht},
  year      = {1981},
  doi       = {10.1007/978-94-015-7658-1_21},
  isbn      = {978-94-015-7658-1}
}

@Article{KRALCHEVSKY2000145,
  author  = {Kralchevsky, P. and Nagayama, K.},
  title   = {Capillary interactions between particles bound to interfaces, liquid films and biomembranes},
  journal = {Advances in Colloid and Interface Science},
  volume  = {85},
  number  = {2},
  pages   = {145--192},
  year    = {2000},
  doi     = {10.1016/S0001-8686(99)00016-0},
  url     = {https://doi.org/10.1016/S0001-8686(99)00016-0}
}

@Book{kralchevsky2001particles,
  author    = {Kralchevsky, P. and Nagayama, K.},
  title     = {Particles at Fluid Interfaces and Membranes: Attachment of Colloid Particles and Proteins to Interfaces and Formation of Two-Dimensional Arrays},
  publisher = {Elsevier Science},
  year      = {2001},
  isbn      = {9780080538471},
  url       = {https://books.google.com.ua/books?id=Lsvn5zl7gvsC}
}

@Article{Gallo2023,
  author  = {Gallo, Mirko and Magaletti, Francesco and Georgoulas, Anastasios and Marengo, Marco and De Coninck, Joel and Casciola, Carlo Massimo},
  title   = {A nanoscale view of the origin of boiling and its dynamics},
  journal = {Nature Communications},
  volume  = {14},
  number  = {1},
  pages   = {6428},
  year    = {2023},
  doi     = {10.1038/s41467-023-41959-3},
  url     = {https://doi.org/10.1038/s41467-023-41959-3},
  issn    = {2041-1723}
}

@Article{Magaletti2021,
  author  = {Magaletti, Francesco and Gallo, Mirko and Casciola, Carlo Massimo},
  title   = {Water cavitation from ambient to high temperatures},
  journal = {Scientific Reports},
  volume  = {11},
  number  = {1},
  pages   = {20801},
  year    = {2021},
  doi     = {10.1038/s41598-021-99863-z},
  url     = {https://doi.org/10.1038/s41598-021-99863-z},
  issn    = {2045-2322}
}

@Article{Wang2016,
  author  = {Wang, Sen and Javadpour, Farzam and Feng, Qihong},
  title   = {Confinement Correction to Mercury Intrusion Capillary Pressure of Shale Nanopores},
  journal = {Scientific Reports},
  volume  = {6},
  number  = {1},
  pages   = {20160},
  year    = {2016},
  doi     = {10.1038/srep20160},
  url     = {https://doi.org/10.1038/srep20160},
  issn    = {2045-2322}
}

@Article{Cao2019,
  author  = {Cao, C. R. and Huang, K. Q. and Shi, J. A. and Zheng, D. N. and Wang, W. H. and Gu, L. and Bai, H. Y.},
  title   = {Liquid-like behaviours of metallic glassy nanoparticles at room temperature},
  journal = {Nature Communications},
  volume  = {10},
  number  = {1},
  pages   = {1966},
  year    = {2019},
  doi     = {10.1038/s41467-019-09895-3},
  url     = {https://doi.org/10.1038/s41467-019-09895-3},
  issn    = {2041-1723}
}

@Article{Hajimirza2021,
  author  = {Hajimirza, Sahand and Gonnermann, Helge M. and Gardner, James E.},
  title   = {Reconciling bubble nucleation in explosive eruptions with geospeedometers},
  journal = {Nature Communications},
  volume  = {12},
  number  = {1},
  pages   = {283},
  year    = {2021},
  doi     = {10.1038/s41467-020-20541-1},
  url     = {https://doi.org/10.1038/s41467-020-20541-1},
  issn    = {2041-1723}
}

@Article{Palodkar2017,
  author  = {Palodkar, Avinash V. and Jana, Amiya K.},
  title   = {Formulating formation mechanism of natural gas hydrates},
  journal = {Scientific Reports},
  volume  = {7},
  number  = {1},
  pages   = {6392},
  year    = {2017},
  doi     = {10.1038/s41598-017-06717-8},
  url     = {https://doi.org/10.1038/s41598-017-06717-8},
  issn    = {2045-2322}
}

@Article{Palodkar2018,
  author  = {Palodkar, Avinash V. and Jana, Amiya K.},
  title   = {Fundamental of swapping phenomena in naturally occurring gas hydrates},
  journal = {Scientific Reports},
  volume  = {8},
  number  = {1},
  pages   = {16563},
  year    = {2018},
  doi     = {10.1038/s41598-018-34926-2},
  url     = {https://doi.org/10.1038/s41598-018-34926-2},
  issn    = {2045-2322}
}

@Article{Petters2020,
  author  = {Petters, Markus and Kasparoglu, Sabin},
  title   = {Predicting the influence of particle size on the glass transition temperature and viscosity of secondary organic material},
  journal = {Scientific Reports},
  volume  = {10},
  number  = {1},
  pages   = {15170},
  year    = {2020},
  doi     = {10.1038/s41598-020-71490-0},
  url     = {https://doi.org/10.1038/s41598-020-71490-0},
  issn    = {2045-2322}
}

@Article{PhysRevE.105.015301,
  author  = {Lulli, Matteo and Biferale, Luca and Falcucci, Giacomo and Sbragaglia, Mauro and Shan, Xiaowen},
  title   = {Mesoscale perspective on the Tolman length},
  journal = {Phys. Rev. E},
  volume  = {105},
  number  = {1},
  pages   = {015301},
  year    = {2022},
  doi     = {10.1103/PhysRevE.105.015301},
  url     = {https://link.aps.org/doi/10.1103/PhysRevE.105.015301}
}

@Article{10.1063/1.469505,
  author  = {Alejandre, Jos{\'e} and Tildesley, Dominic J. and Chapela, Gustavo A.},
  title   = {Molecular dynamics simulation of the orthobaric densities and surface tension of water},
  journal = {The Journal of Chemical Physics},
  volume  = {102},
  number  = {11},
  pages   = {4574--4583},
  year    = {1995},
  doi     = {10.1063/1.469505},
  url     = {https://doi.org/10.1063/1.469505},
  issn    = {0021-9606}
}

@Article{Burian2024,
  author  = {Burian, Sergii and Shportun, Yevhenii and Yaroshchuk, Andriy and Bulavin, Leonid and Lacroix, David and Isaiev, Mykola},
  title   = {Size-Dependent Wetting Contact Angles at the Nanoscale Defined by Equimolar Surfaces and Surfaces of Tension},
  journal = {Scientific Reports},
  volume  = {14},
  number  = {1},
  pages   = {31340},
  year    = {2024},
  doi     = {10.1038/s41598-024-82683-2},
  url     = {https://doi.org/10.1038/s41598-024-82683-2},
  issn    = {2045-2322}
}

@Article{Azouzi2013,
  author  = {Azouzi, Mouna El Mekki and Ramboz, Claire and Lenain, Jean-Fran{\c c}ois and Caupin, Fr{\'e}d{\'e}ric},
  title   = {A coherent picture of water at extreme negative pressure},
  journal = {Nature Physics},
  volume  = {9},
  number  = {1},
  pages   = {38--41},
  year    = {2013},
  doi     = {10.1038/nphys2475},
  url     = {https://doi.org/10.1038/nphys2475},
  issn    = {1745-2481}
}

@Article{Cui_2021,
  author    = {Shu-Wen Cui and Jiu-An Wei and Qiang Li and Wei-Wei Liu and Ping Qian and Xiao Song Wang},
  title     = {Tolman length of simple droplet: Theoretical study and molecular dynamics simulation},
  journal   = {Chinese Physics B},
  year      = {2021},
  month     = {jan},
  publisher = {Chinese Physical Society and IOP Publishing Ltd},
  volume    = {30},
  number    = {1},
  pages     = {016801},
  doi       = {10.1088/1674-1056/abb65a},
  url       = {https://dx.doi.org/10.1088/1674-1056/abb65a}
}

@Article{10.1063/1.3493464,
  author  = {Block, Benjamin J. and Das, Subir K. and Oettel, Martin and Virnau, Peter and Binder, Kurt},
  title   = {Curvature dependence of surface free energy of liquid drops and bubbles: A simulation study},
  journal = {The Journal of Chemical Physics},
  volume  = {133},
  number  = {15},
  pages   = {154702},
  year    = {2010},
  month   = {10},
  issn    = {0021-9606},
  doi     = {10.1063/1.3493464},
  url     = {https://doi.org/10.1063/1.3493464}
}

@Article{Joswiak2013,
  author  = {Joswiak, Mark N. and Duff, Nathan and Doherty, Michael F. and Peters, Baron},
  title   = {Size-Dependent Surface Free Energy and Tolman-Corrected Droplet Nucleation of TIP4P/2005 Water},
  journal = {The Journal of Physical Chemistry Letters},
  volume  = {4},
  number  = {24},
  pages   = {4267--4272},
  year    = {2013},
  doi     = {10.1021/jz402226p},
  url     = {https://doi.org/10.1021/jz402226p}
}

@article{Joswiak2016,
  author  = {Joswiak, Mark N. and Do, Ryan and Doherty, Michael F. and Peters, Baron},
  title   = {Energetic and entropic components of the Tolman length for mW and TIP4P/2005 water nanodroplets},
  journal = {The Journal of Chemical Physics},
  volume  = {145},
  number  = {20},
  pages   = {204703},
  year    = {2016},
  month   = {11},
  doi     = {10.1063/1.4967875},
  url     = {https://doi.org/10.1063/1.4967875}
}

@article{Wilhelmsen2015,
  author  = {Wilhelmsen, {\O}ivind and Bedeaux, Dick and Reguera, David},
  title   = {Tolman length and rigidity constants of the Lennard-Jones fluid},
  journal = {The Journal of Chemical Physics},
  volume  = {142},
  number  = {6},
  pages   = {064706},
  year    = {2015},
  month   = {02},
  doi     = {10.1063/1.4907588},
  url     = {https://doi.org/10.1063/1.4907588}
}

@article{Kanduc2017,
  author  = {Kanduc, Matej},
  title   = {Going beyond the standard line tension: Size-dependent contact angles of water nanodroplets},
  journal = {The Journal of Chemical Physics},
  volume  = {147},
  number  = {17},
  pages   = {174701},
  year    = {2017},
  month   = {11},
  doi     = {10.1063/1.4990741},
  url     = {https://doi.org/10.1063/1.4990741}
}
\end{document}